\documentclass[fleqn,usenatbib]{mnras}
\pdfoutput=1
\usepackage{newtxtext,newtxmath}

\usepackage[T1]{fontenc}
\usepackage{array}

\DeclareRobustCommand{\VAN}[3]{#2}
\let\VANthebibliography\thebibliography
\def\thebibliography{\DeclareRobustCommand{\VAN}[3]{##3}\VANthebibliography}

\usepackage{graphicx}	
\graphicspath{{./figures/}} 
\usepackage{amsmath}	
\usepackage{placeins}

\title[Holistic Review]{A Holistic Review of a Galactic Interaction}

\author[D. Grion Filho et al.]{
Douglas Grion Filho$^{1},$\thanks{E-mail: df2629@columbia.edu}
Kathryn V. Johnston,$^{1,2}$
Eloisa Poggio,$^{3,4}$
Chervin F. P. Laporte,$^{5,6,7,4}$
\newauthor
Ronald Drimmel,$^{3}$ 
and Elena D'Onghia$^{8}$
\\
$^{1}$Department of Astronomy, Columbia University, 550 West 120th Street, New York, NY 10027, USA \\
$^{2}$Center for Computational Astrophysics, Flatiron Institute, 162 5th Av., New York City, NY 10010, USA\\
$^{3}$Osservatorio Astrofisico di Torino, Istituto Nazionale di Astrofisica (INAF), I-10025 Pino Torinese, Italy\\
$^{4}$Université Côte d’Azur, Observatoire de la Côte d’Azur, CNRS, Laboratoire Lagrange, France\\
$^{5}$ Institut de Ci\`encies del Cosmos (ICCUB), Universitat de Barcelona (IEEC-UB), Mart\'i i Franqu\`es 1, 08028 Barcelona, Spain\\
$^{6}$Kavli Institute for the Physics and Mathematics of the Universe (WPI), The University of Tokyo Institutes for Advanced Study (UTIAS)\\
The University of Tokyo, Chiba 277-8583, Japan\\
$^{7}$Department of Physics \& Astronomy, University of Victoria, 3800 Finnerty Road, Victoria BC, V8P 5C2 Canada\\
$^{8}$University of Wisconsin, Madison, Astronomy Department, 475 N Charter Str., Wisconsin, USA
}

\date{Accepted XXX. Received YYY; in original form ZZZ}

\pubyear{2020}

\begin{document}
\label{firstpage}
\pagerange{\pageref{firstpage}--\pageref{lastpage}}
\maketitle

\begin{abstract}

Our situation as occupants of the Milky Way Galaxy, bombarded by the Sagittarius dwarf galaxy, provides an intimate view of physical processes that can lead to the dynamical heating of a galactic disk. 
While this evolution is instigated by Sagittarius, it is also driven by the intertwined influences of the dark matter halo and the disk itself.
We analyse an N-body simulation following a Sagittarius-like galaxy interacting with a Milky-Way-like host to disentangle these different influences during the stages of a minor merger. 
The accelerations in the disk plane from each component are calculated for each snapshot in the simulation, and then decomposed into Fourier series on annuli.
The analysis maps, quantifies and compares the scales of the individual contributions over space and through time:
(i) accelerations due to the satellite are only important around disk passages;
(ii) the influence around these passages is enhanced and extended by the distortion of the dark matter halo;
(iii) the interaction drives disk asymmetries within and perpendicular to the plane and the self-gravity of these distortions increase in importance with time eventually leading to the formation of a bar.
These results have interesting implications for identifying different influences within our own Galaxy. 
Currently, Sagittarius is close enough to a plane crossing to search for localized signatures of its effect at intermediate radii, the distortion of the Milky Way's dark matter halo should leave its imprint in the outer disk and the disk’s own self-consistent response is sculpting the intermediate and inner disk.

\end{abstract}


\section{Introduction}

Recently, there has been a significant revision in our perception of the Milky Way (hereafter MW), shifting from an equilibrium picture to one rife with signatures of disequilibrium, as revealed by large-scale stellar surveys.

North/South asymmetries in density and velocity structure were found in the Galactic disk using data from SDSS \cite{widrow12,williams13}. Analysis of LAMOST and RAVE suggested these local asymmetries to be part of a large-scale vertical wave \citep{carlin13,xu15}. Studies motivated by observations of M-giants selected using 2MASS photometry showed this wave to be propagating from the Solar Neighborhood to the outer disk \citep{price-whelan15,li17,sheffield18,bergemann18}. In addition to this, the disk of our Galaxy presents a large-scale warp \citep[][]{burke57,kerr57, djorgovski89, lopezcorredoira2002,reyle2009,amores2017,chen2019,skowron2019}, suggesting that the Milky Way might be responding to external/internal torques and perturbations. Indeed, the initial discovery of disk asymmetries in SDSS led \cite{widrow14} to investigate the nature of "bending" and "breathing" modes in galactic disks  \citep[see][for an early discussion of bending modes]{dubinski95} and there have been several studies of their appearance in N-body and cosmological hydrodynamical simulations of interactions between satellite and parent galaxies \citep{gomez13, donghia16, gomez17, laporte18a}. 
The second data release from the {\it Gaia} mission (hereafter {\it Gaia} DR2) connected the disparate lines of evidence from prior surveys with a global view that mapped these ripples in the disk on a "macroscopic" scale \citep{gaia18,friske19} in both vertical and radial directions \citep{antoja18}. Vertical motions also suggest that the Galactic warp might be evolving with time \citep{Poggio2020a,Cheng2020}. The "microscopic" counterpart of the ripples was revealed by {\it Gaia} DR2 in the form of spirals in the vertical component of phase-space (e.g. in density the $z$-$v_z$ plane) - a signature of incomplete phase-mixing following a disturbance \citep{antoja18,binney18,laporte19,bland-hawthorn19}. 


In the plane of the disk, several kinematic ridges in the $U-V$ velocity plane have been revealed in the Solar neighborhood \citep{gaia18b,ramos18}. Other ridges have also been shown in the ($R$, $v_T$) plane \citep{antoja18, kawata18, laporte20b} and in action space \citep{trick19}.  Furthermore, \citet{gaia18b, friske19, eilers20} have shown both local and global spiral waves in mean galactocentric radial velocities. Several dynamical processes could explain their origin, such as spiral arms \citep{quillen03, quillen05, sellwood19, khoperskov20, hunt19}, bar resonances \citep{fragkoudi19, fragkoudi20, hunt19, laporte20b,kawata20, trick21, monari19}, or the response of the disc to an infalling satellite \citep{minchev09, gomez12, antoja18, laporte19, khanna19}.

Possible culprits for causing vertical and in-plane disturbances are the satellite galaxies that orbit the MW. (Note that \cite{khoperskov19} pointed to buckling bar as another plausible explanation for vertical disturbances.) 
Parent/satellite galaxy interactions enjoy a rich literature around the topics of dynamical friction, satellite disruption and disk response \citep{johnston95,walker96}. 
Much of this past work has concentrated on the long-term and orbit-averaged effect of dynamical heating of disks, and the implications for galaxy evolution \citep{toth92,velazquez99}. In contrast, the detail of the current Milky Way data sets instead offers the opportunity to study the mechanisms of disk heating in process. 

Past work has also emphasized and explored the importance of all galactic components --- disk, satellite and dark matter halo (hereafter DM halo) --- in driving dynamical evolution through mutual gravitational interactions
\citep{weinberg98,weinberg99,vesperini01, weinberg06,gomez16, laporte18a, laporte18b, garavito20a}: 
the satellite deforms the DM halo; the deformation of the DM halo both affects the orbit of the satellite and can in turn perturb the disk; the perturbation of the disk can drag on both the satellite and the DM halo; and the satellite can directly affect the disk during pericenters and plane crossings.
Recently, \citet{laporte18b} used N-body simulations to demonstrate that the interaction of the MW with its closest satellite companion, the Sagittarius dwarf galaxy (herefter Sgr) could plausibly produce many of the disk disturbances that had been noted prior to {\it Gaia} DR2, both on large scales and small scales. They confirmed the two-phase seeding mechanism for the disk response through the action of torques from the dark matter halo wake later transitioning to tides from Sgr itself. These same models were subsequently found to contain similar scales in the detailed phase-structure in the {\it Gaia} data set in multiple dimensions \citep{laporte19}. Subsequent work \citep{bland-hawthorn20, hunt21} connects these features with even higher resolution simulations.

\begin{figure}
    \centering
    \includegraphics[width=7.cm]{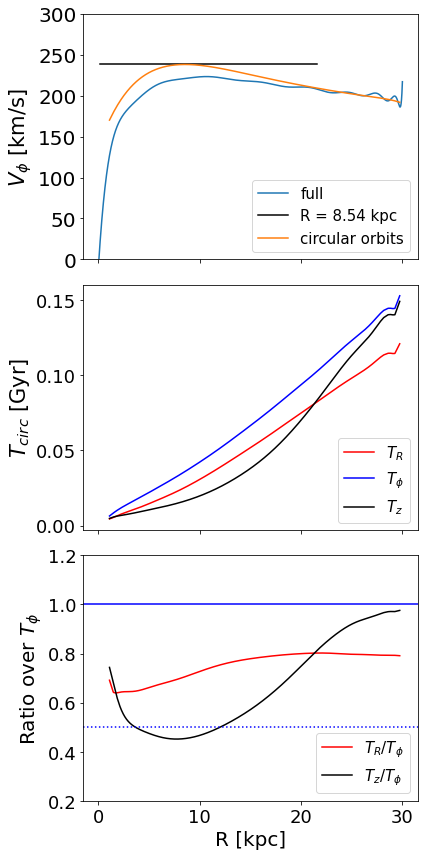}
    \caption{Rotation curve of the disk for the first snapshot (top). Orange line is the mean rotation curve for particles on roughly circular orbits, while the blue line is mean $V_{\phi}$ for all particles. The black line at $R = 8.54$ kpc indicates the point at which $dV_{\rm circ}/dt = 0$. The middle panel shows the periods associated with the three fundamental frequencies ($\kappa$, $\Omega$, and $\nu$ respectively) for particles on circular orbits. Finally, the bottom panel shows the ratio of these periods with respect to $T_{\phi}$, with a dotted line at a ratio equal to 0.5 and a solid line at 1.0.}
    \label{fig:rot_curve}
\end{figure}

\begin{figure*}
    \centering
    \includegraphics[width=13.cm]{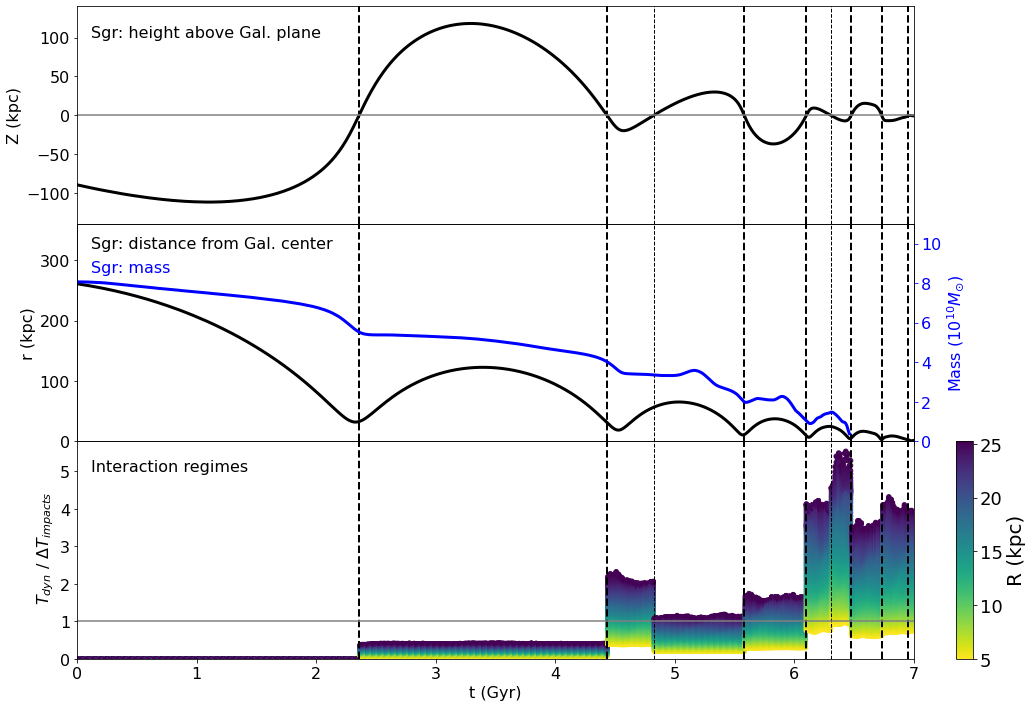}
    \caption{Interaction regimes (reproduced from Figure 1 of \citet{poggio20}). Top panels show height above the Galactic plane ($z$, black line, upper panel), distance from the Galactic center in spherical coordinates ($r$, black line, middle panel) and mass loss as a function of time (blue line, middle panel) of the Sagittarius Dwarf Galaxy as a function of time. The mass-loss curve ends at the approximate "present day" snapshot (6.4 Gyr). Vertical lines represent the times of disk crossings. The bottom panel demonstrates how the simulation can be broadly split into regimes by comparing the local orbital time $T_{\rm dyn}$ to the time between impacts $\Delta T_{\rm impact}$: in Regime 0 before the first impact the disk evolves in isolation; in Regime 1 between impacts 1 and 2 where the effect of a single impact can be studied; in Regime 2 the disk is being impacted repeatedly on similar timescales to an orbit; and in Regime 3 the impacts occur multiple times per orbit.}
    \label{fig:orbit}
\end{figure*}

This paper characterizes the complex dynamical interaction between the Milky Way's disk,  DM halo and Sgr. Our aim is to retrieve what spatial and temporal regimes in the disk response can be attributed to the influence of each source of perturbation. 
To this end, we work with one specific model from the series of N-body simulations developed in \cite{laporte18b} that best reflects the scale of the response seen in the Milky Way.
Our paper is a companion to the analysis of \citet{poggio20}, who used Fourier Series to interpret the vertical evolution of the disk in this same simulation \citep[see also][for prior work also using Fourier decomposition to study disk evolution]{chequers17, chequers18, laporte18b}.
We add to this study by first looking at Fourier Series of the response of the disk in density as a Fourier series over time at each radius. We subsequently decompose the {\it acceleration} due to each component (disk, DM halo, and Sgr) in the disk plane.
This approach allows us to compare amplitude, spatial scales and timescales of these different influences and connect responses to their origin as a natural follow-up to \citep{laporte18b}.

The paper is structured as follows: Section 2 goes over the simulations and numerical methods used in this work;  
Section 3  explores the context for our study of influences by first reviewing the evolving properties of Sgr and the disk's density structure; Section 4 shows the results of our Fourier analysis of drivers of disk evolution; and
Section 5 summarizes our results and presents our conclusions.

\section{Numerical Methods}

\subsection{Simulation Description}
The simulation used in this work is taken from \citet{laporte18b}, which report on a series of live N-body simulations of the interaction between a Sgr-like galaxy with a MW-like host. The work contains several different models, some of them also taking into account the effects of the Large Magellanic Cloud (LMC) on a first infall orbit. In this work, however, we focus on the L2 model of \citet{laporte18b}, which looks only at the interaction with Sgr alone. We choose the L2 model because other more massive progenitor models produced disk responses with amplitude in excess of what is observed \citep[see discussion in][]{laporte19}. Unlike prior models, the simulation follows the response of the Galaxy since the time of first crossing of the virial radius to the present-day, which was central to depict and characterise the early influence of the dark matter halo wake \citep[see also][]{weinberg98, vesperini01, gomez16} excited by Sgr on the disk during the course of its evolution since z~1.

The N-body simulation presented here ran with the tree-code $GADGET-3$. The simulated MW had the following properties: a spherical Dark Matter Halo following a Hernquist profile \citep{hernquist90} with mass $M_h = 10^{12} M_{\odot}$ and a scale length $a_h = 52$ kpc, an exponential disk of $M_{\rm disk} = 6 \times 10^{10} M_{\odot}$, with a scale length of $R_d = 3.5\ kpc$, scale height of $h_d = 0.53$ kpc and a central bulge represented by a Hernquist Sphere with mass of $M_{\rm bulge} = 10^{10} M_{\odot}$ and scale length $a = 0.7$ kpc. 
The DM halo is adiabatically contracted leading to a final  circular  velocity  of $V_{\rm circ} =  239$ km/s at $R_0= 8$ kpc as shown by Figure \ref{fig:rot_curve}. 

Sgr's dark matter halo in these simulations is parametrized using a Hernquist profile with varying masses and scale lengths depending on the model. For the model we are using in this analysis - the L2 model - Sgr is given a virial mass model of $M_{200} = 6.0 \times 10^{10} M_{\odot}$, and its Hernquist parameters are $M_h = 8.0 \times 10^{10} M_{\odot}$ and $a_h = 8$ kpc. These parameters for Sgr are chosen  to  closely  match  their  corresponding NFW halos but with steeper fall-off  at  large  radii. The stellar content matches the stellar-halo mass relation from abundance matching or galaxy formation models \citep{moster13,sawala16}.
The final dwarf properties  are broadly consistent with recent observational work on the progenitor stellar mass and stream velocity dispersions \citep[e.g.][]{niederste-ostholt10,gibbons17}.

The  particle masses in the simulation are $m_h = 2.6 \times 10^{4} M_{\odot}$, $m_d = 1.2 \times 10^{4} M_{\odot}$, and $m_b = 10^{4} M_{\odot}$ for the dark matter, disk and bulge components respectively. The softening lengths for the disk are $\epsilon_h = 60$ pc, $\epsilon_d = \epsilon_b = 30$ pc. The simulation represents the disk with 5 million particles and the DM halo with 40 million particles, such as to minimize the effect of heating between the different species.

Our snapshot analyses are performed in a reference frame centered on the bulge and aligned with the galactic midplane, as outlined in \cite{gomez13} and \cite{laporte18b}. This means that our analysis does not follow the re-orientation or motion  of the disk relative to initial conditions, but instead looks at the net acceleration the disk experiences relative to its own geometrical configuration.

\subsection{Characterizing influences on the disk}

For each snapshot in the simulation we assess the influence that the different components have on the disk by binning the disk midplane in radius and azimuth using a 100x100 polar grid -- i.e. dividing $R$ into 100 slices ranging from 0-30 kpc, while also slicing $\phi$ into 100 slices from 0 - 2$\pi$. We then estimate the acceleration at each grid point, $i$ due to each component (disk, DM halo, and Sgr) by simply summing over individual forces from each particle, 
\begin{equation}
    a_{q,i} = \sum_{j=0}^{N_{\rm comp}} - \frac{G m_{j}}{(d_{ij} + \alpha)^2 d_{ij}} \times (q_{i} - q_{j})
    \label{eqn:acceleration}
\end{equation}
where $d$ is the distance between the component particle and the $z = 0$ grid point, $q$ is the coordinate ($x$, $y$ or $z$) and $\alpha = 0.7$ kpc is a softening parameter to lessen the effects of component particles that just so happen to lie in the $z = 0$ plane. 
Since we are interested in drivers of the evolution of the disk's structure rather than its overall motion we assess the tidal accelerations from  Sgr and the DM Halo by subtracting the acceleration at the disks center at each grid point.  

\begin{figure}
    \centering
    \includegraphics[width=7.cm]{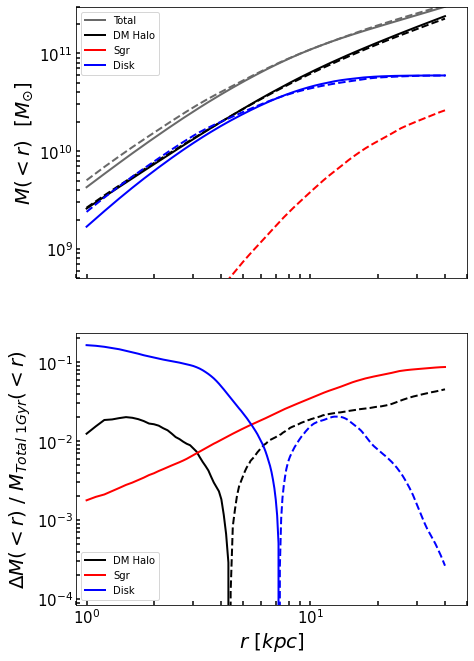}
    \caption{Mass evolution as an indicator of the scale of overall net acceleration contributions ($\sim G M(<r)/r^2$) and changes to those contributions. \textit{Top}: Total mass $M(<r)$ enclosed within spherical radius $r$ (gray), as well as separate contributions from the DM halo, Sgr, and Disk. For the separate components the solid line indicates $M(<r)$ at $t=1$Gyr  and the dotted line is for the end of the simulation. \textit{Bottom}: Change between dotted and solid lines from top panel divided by the total $M(<r)$ at $t=$1Gyr (bold grey line from top panel). Solid lines indicate positive $\Delta M(<r)$ while dashed lines indicates negative $\Delta M(<r)$. Note the small evolution in the radial profiles, and even smaller contribution of Sgr debris to that change.}
    \label{fig:mass distribution}
\end{figure}

\begin{figure*}
    \centering
    \includegraphics[width=18.5cm]{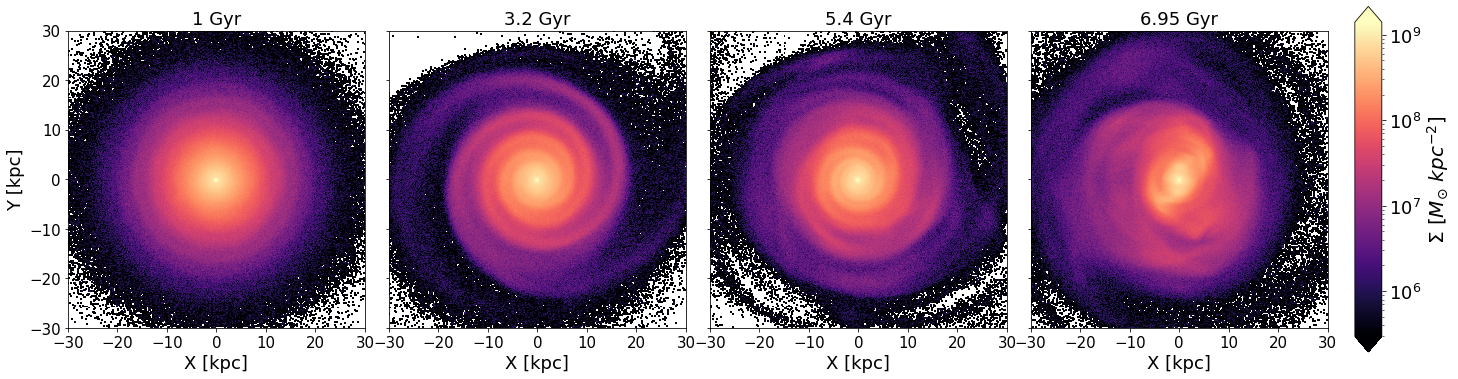}
    \caption{Panels each show an example of the disk's surface density from each Epoch in the simulation. They have has separate morphological characteristics, going from isolation to spiral arm formation, to a phase of winding spiral arms, and finally mostly complex structure with the bar forming in the inner disk.}
    \label{fig:regimes}
\end{figure*}


We utilize Fourier Transforms in order to summarize both the response (in density) of and influences (by projecting our acceleration fields into cylindrical coordinates) on the disk. 
For a given ring at Galactocentric radius $R$, we express deviations from initial conditions $\Delta Q$ in the quantity $Q$  as an infinite sum of Fourier terms in Galactic azimuth $\phi$:
\begin{equation}
    \Delta Q(R,\phi) = A_0(R) + \sum_{m=1}^{\infty} A_m(R) \cos(m \phi - \phi_m(R))
    \label{eqn:Fourier}
\end{equation}
where $A_m(R)$ is the amplitude of the $m$th Fourier term and $\phi_m(R)$ is the phase. 
(Note that the only non-zero component in any case early in the simulation will be the radial acceleration, $a_R$, which is keeping the disk particles in orbit.)
The azimuthal angle $\phi$ is taken as positive in the direction of Galactic rotation. Rings in $R$ are chosen to contain the same number of particles, so they are narrower in the inner and wider in the outer disk.

\section{Results I: Overview of the Interaction and Responses}
\label{sec:context}

This section first reviews the {\it responses} to the intertwined gravitational interactions by summarizing the evolution in the simulation, in particular for Sgr (Section \ref{sec:sgr}) and the disk  (Sections \ref{sec:response} and \ref{sec:disk}). We then go on to distinguish {\it drivers} of that evolution in Section \ref{sec:results}.

\subsection{Sagittarius' mass and orbit, interaction regimes and epochs, and the global mass profile}

\label{sec:sgr}

In this simulation, Sagittarius (Sgr) is the instigator of the evolution of the disk. Figure \ref{fig:orbit}, taken from \cite{poggio20}, shows the orbit of the satellite both in spherical radius, $r$ and vertical position, $z$, along with an estimate for its instantaneous mass (described in Carr et al, {\it in prep}). We can broadly split the simulation into into four "epochs" in time.
\begin{itemize}
    \item {Epoch 0}: Prior to the first impact we can examine how the disk evolves in isolation. We will use the scale of Fourier amplitudes in this regime to understand the significance of those seen in all subsequent regimes. \citep[see Appendix C of][which demonstrates that the nature of features seen in this regime are in common with those strictly in isolation.]{poggio20}
    \item {Epoch 1}: Between impacts 1 and 2 interactions are far enough apart to allow the disk to relax between plane-crossings and we can study this secular, internal process.
    \item {Epoch 2}: Impacts 2-5 cross the plane on timescales comparable to the orbital timescales in the middle disk so we might expect some resonant responses. 
    \item {Epoch 3}: For the remainder of the simulation repeated impacts on short time-scales cause a strong and complicated response in the Galactic disk.
\end{itemize}

The bottom panel of Figure \ref{fig:orbit} shows how different regions of the disk  --- e.g. inner ($<$ 5 kpc, yellow) , middle (5-15 kpc, green), and outer ($>$ 15 kpc, blue)  ---  are directly impacted in different ways during each epoch and expected to respond on different timescales. 
Following \citet{poggio20}, we adopt 4 characteristic interaction "Regimes" in space (which are time-dependent, based on the ratio $ T_{\rm dyn} / \Delta T_{\rm impacts}$, where $ \Delta T_{\rm impacts}$ is the time interval between two impacts of Sgr and $ T_{\rm dyn}$ is the local orbital time at cylindrical radius $R$ in the disk plane ($ \sim 2 \pi R / V_{\phi}$, where $ V_{\phi}$ is the mean azimuthal velocity). 
\begin{itemize}
\item Regimes, 0, 1, 2, and 3 correspond to a ratio of $ T_{\rm{dyn}} / \Delta T_{\rm{impacts}}$ of 0, less than unity, of order unity and greater than unity respectively, i.e. the disk evolving in isolation [0], or  in response to: a single impact [1]; impacts repeated roughly every orbit [2], or multiple impacts per orbit [3].
\end{itemize}

Figure \ref{fig:mass distribution} (top) gives a global view of the overall evolution by showing how the enclosed mass within a spherical radius $r$ --- $M(<r)$ --- changes from the beginning (1 Gyr --- solid lines) to the end of the simulation (6.95 Gyr --- dashed lines) for each component in the simulation: Sgr (red), DM halo (black) and disk (blue). Note that this is also an indicator of changes to the net acceleration at each radius. The bottom panel of the figure shows the fractional change $\Delta M(<r)/M(r)$ relative to the initial  DM halo distribution  for each component. The figure demonstrates that the mass distribution of the DM halo does not significantly change throughout the simulation, though there is a small inward migration of mass into the disk regions. The disk experiences most change in its mass distribution within the inner disk ($r < 5$ kpc). Sgr's debris contributes a negligible fraction of these mass changes within the disk.  The plot  emphasizes that the overall mass redistribution, though measurable, is small (order percent changes). 

\begin{figure*}
    \centering
    \includegraphics[width=17.5cm]{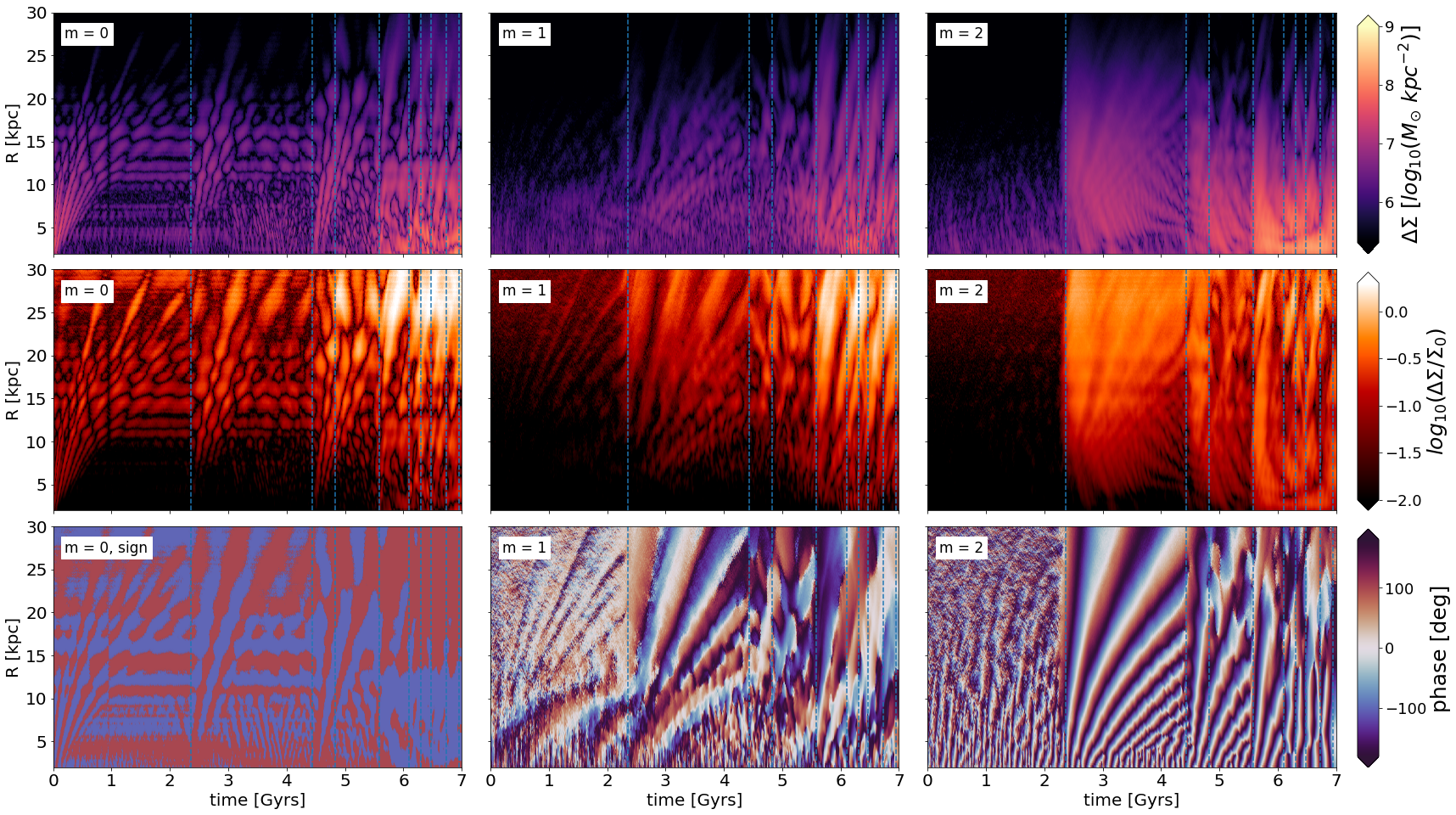}
    \caption{Amplitudes and phases respectively of the spatial Fourier terms for the m=0 (first column), m=1 (second column), and m=2 (third column) terms in changes to the disk surface density ($\Delta \Sigma$). The amplitude for the m=0 term is colored as the difference in amplitude from the snapshot at 1Gyr, which we define as the isolation snapshot ($\Delta\Sigma = \Sigma\ - \Sigma_0$). Rows are then organized as density (top), percentage change from the isolation snapshot ($\Delta\Sigma/\Sigma_0$, middle), and phases (bottom panel). Vertical dashed lines represent impact events with Sgr, and the "phase" in the bottom left-hand panel indicates the sign of the $m=0$ amplitude.}
    \label{fig:Fourier density}
\end{figure*}

\subsection{Disk density evolution}

\label{sec:response}

Figure \ref{fig:regimes} gives a  visual idea of the disk response by showing a sample snapshot of disk surface density in each Epoch (0 to 3 in sequence in left to right panels).
Since there is a strong radial dependence for each spatial Regime (bottom of Figure \ref{fig:orbit}), some can co-exist at different radii within each panel. 
For example, the last panel of figure \ref{fig:regimes}, showing the disk at time $t = 6.95$ Gyr contains Regime 1,2, and 3 depending on whether one is considering the inner, middle or outer regions of the disk.  
In Epoch 0 the disk is stable and almost featureless, {other than  some low-level self-consistent "ringing" left over from the initialization routine (see below for more discussion)}; in Epoch 1 the first impact drives the disk to form a regular 2-armed spiral pattern; in Epoch 2 the repeated impacts on orbital times excite multiple overlapping spiral patterns; in Epoch 3 the impacts themselves are the dominant influence and the disk is sufficiently perturbed to form a bar.




Figure \ref{fig:Fourier density} summarizes the evolution of the disk by presenting Fourier decompositions of deviations in the disk surface density, $\Delta \Sigma$, from its state at $t_0=1$Gyr.
This allows us to show a quantitative representation of the evolution throughout the entire simulation over space (i.e. radius on the $y$-axis) and for every snapshot in time (on the $x$-axis), with six numbers --- i.e. the amplitudes and phases of the $m=0,1,2$ Fourier terms in the left/middle/right columns respectively.
The rows show: (top) the absolute change in $\Sigma$, with the same color bar as Figure \ref{fig:regimes} for comparison; (middle) the percent change relative to the unperturbed $m=0$ term; and (bottom) the phase. Note that for the $m=0$ term the "phase" simply indicates whether the change is positive (red) or negative (blue) since this component is, by definition, axisymmetric. 
In order to account for the numerical scatter, we smooth the amplitude values at each radius, $R$, using a simple moving median over a 0.12 Gyr window.
The vertical line in all the panels indicate the times at which Sgr crosses the disk plane. 

We use Epoch 0, before Sgr's first disk crossing at $\sim$ 2.5 Gyrs,  to characterize the natural, self-gravitating response of the disk component to infinitesimal perturbations within the combined disk/halo system. In the real Milky Way these could be due to (e.g.) Giant Molecular Clouds or dark matter subhalos. In the simulations we can see "ringing" in the $m=0$ term left over from the initial conditions, with enhancements (red in bottom left-hand plot) and deficits (blue) in $\Sigma$ propagating outward, with oscillation periods that increase with radius. There is also a response in the $m=1$ term, which is clear in amplitude in the outer disk and steadily oscillating phases at all radii. It is intriguing to note that the phases follow a blue-black-red-white sequence in the inner disk (indicating retrograde pattern rotation) but red-black-blue-white sequence in the outer disk (indicating prograde pattern rotation). The change from retrograde to prograde occurs around $R \sim 10-15$ kpc, where the rotation curve is flat, most dominated by the disk self-gravity and transitioning between the inner rigid body and the outer gently-falling behavior as it becomes halo-dominated. The $m=2$ term shows hints of oscillating phase within $\sim $5kpc, but the amplitude associated with that oscillation is too low for a conclusive identification to be made.

\begin{figure}
    \centering
    \includegraphics[width=8.5cm]{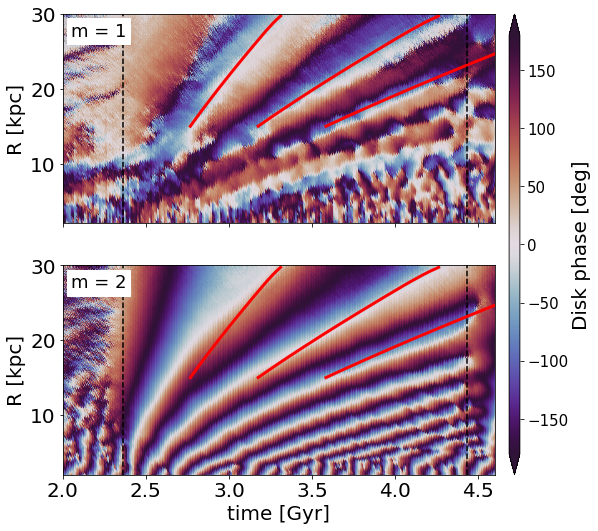}
    \caption{\textit{Top:} Zoom-in of the m=1 phase of the Disk during Epoch 1 (2.3 - 4.4 Gyrs). Over-plotted as a red line starting at $R = 15$kpc is the prediction for the phase given by the rotation curve presented in Figure \ref{fig:rot_curve}. Black vertical lines represent Sgr passages. \textit{Bottom:} Same as top panel but for the m=2 phase.}
    \label{fig:red lines}
\end{figure}

During subsequent Epochs, Figure \ref{fig:Fourier density} shows some patterns that are common to all Fourier terms. 
The influence of Sgr's disk crossings  is immediately apparent in all panels as an increase in amplitude and re-alignment in phase (vertical color-bands coincident with the dotted lines). The only exception is the third disk crossing at $\sim$4.5 Gyrs which occurred at Sgr's orbital apocenter, well beyond the disk extent.
After each crossing, the longer dynamical timescales and slower natural oscillations of the outer disk can be seen in both amplitude and phase as diagonal ridges tipping further and further away from vertical alignment until they are reset by the next interaction. 
Within $R \sim$5kpc, where the rotation curve is closer to rigid body and the axisymmetric bulge has some influence, the oscillation amplitudes are smaller and there is a tendency to maintain alignment in phase.

On the other hand, some features of evolution are unique to each term.

For the \underline{$m=0$ term}, the timescale for oscillation  is shorter than for the higher-order terms (notice that the blue ridges are more closely spaced in time in the bottom left-hand panel compared to the bottom middle and right panels), as a reflection of the higher frequencies in the radial and vertical directions (governing axisymmetric ringing in density) compared to the differences in frequencies in the azimuthal direction (governing winding of non-axisymmetric patterns --- summarized in bottom panel of Figure \ref{fig:rot_curve}). The $m=0$ term also shows the net migration of material around the disk (see Figure \ref{fig:mass distribution}) as it is dominated by a decrease/increase in the disk surface density in the inner/outer disk (blue/red in phase) after $\sim$5.5 Gyrs.
\\ The \underline{$m=1$ term} reveals the same retrograde intrinsic mode in the inner disk seen during Epoch 0. It is replaced by a global prograde pattern for a short period following the second disk crossing at $\sim$2.5 Gyrs, but subsequently resumes. The prograde pattern incited by the disk crossing remains dominant in the outer disk during Epoch 1, but has a lower oscillation frequency than that seen during Epoch 0, and is therefore likely of a different nature. 
The red curves in Figure \ref{fig:red lines}, which follow:
\begin{equation}
    \phi(R,t) = \phi_{\rm cross} + \frac{v_{\rm circ}(R)}{R} (t- t_{\rm cross}),
\end{equation}
where $t_{\rm cross}$ and $\phi_{\rm cross}$ are the time and phase angle (the same for all radii) at disk crossing,
suggest that  the phase evolution is consistent with a local disturbance simply winding up due to the differences of azimuthal time periods with radius, so it can be attributed to phase-mixing.
\\ The \underline{$m=2$ term}, in contrast to the dual nature of the $m=1$ response,  shows changes that are initially aligned in phase at all radii, winding up in smoothly and continuously at all radii to form spiral arms. The red curves in the bottom panel of Figure \ref{fig:red lines} (which are repeated from the top panel) show that the winding up of this pattern is slightly slower than that expected from phase-mixing alone, demonstrating the importance of self-gravity in sustaining the spiral arms.
The power in this term increases dramatically at the center of the disk towards the end of the simulation, with the aligned phases reflecting the presence of a steadily rotating bar (e.g. around 6 Gyrs).

The behavior in the $m=2$ term seen in our simulations is reminiscent of earlier work by \citet{purcell2011} which pointed out how Sgr might instigate the formation of the Milky Way's spiral pattern. As shown by \cite{kyziropoulos2016}, this conclusion about the arms being externally excited rather than plausibly arising from purely internal, secular effects, depends on the combined properties of Sgr and the parent. In our simulations, each disk passage plainly seeds a new set of $\Sigma$ perturbations, so the satellite is clearly responsible.

\begin{figure*}
    \centering
    \includegraphics[width=17.5cm]{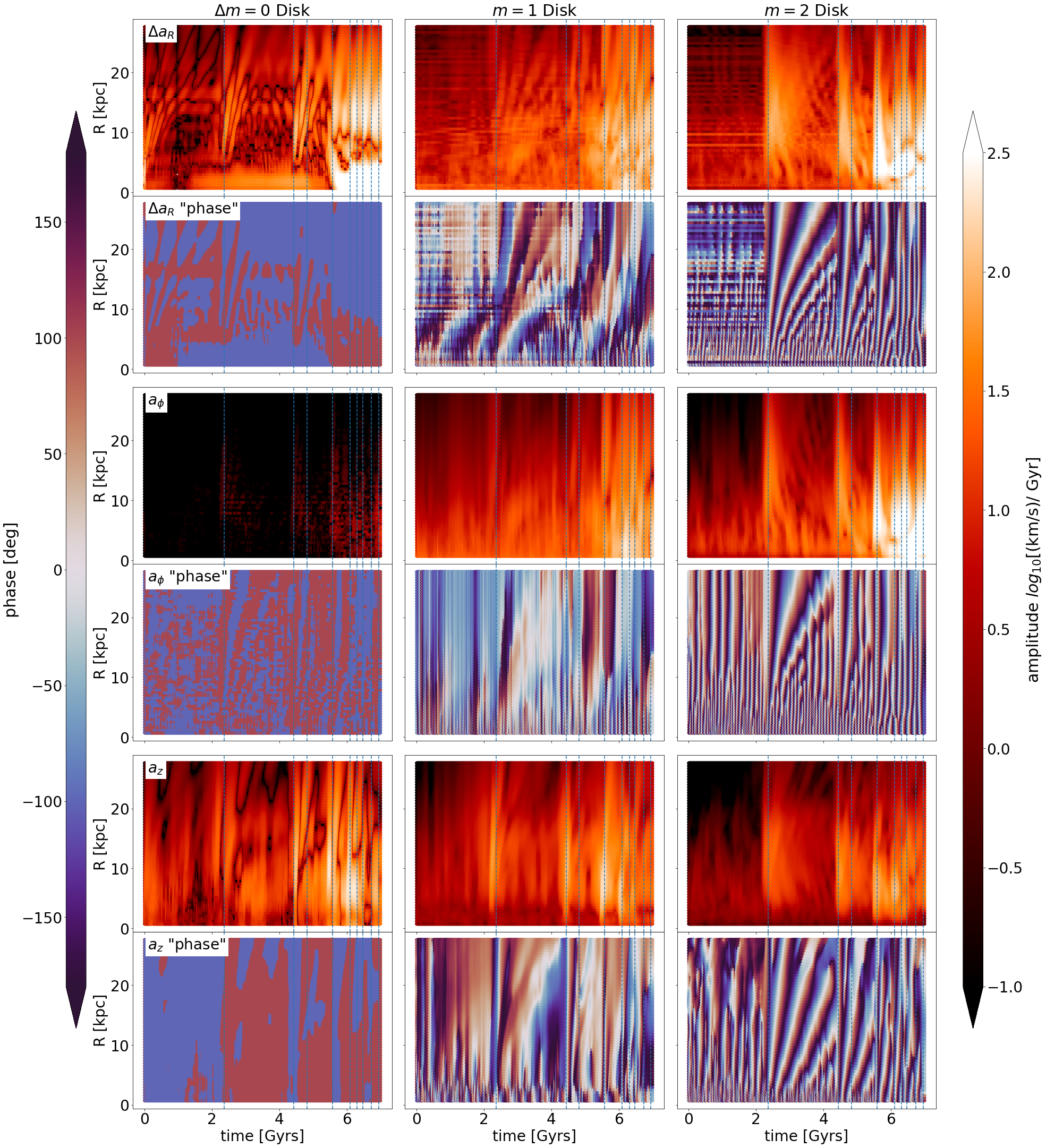}
    \caption{Amplitudes and phases for accelerations due to disk self-gravity. Columns are $\Delta m=0$, $m=1$, and $m=2$, respectively. Rows are the amplitude/phase pairs for $\Delta a_R$, $a_{\phi}$, and $a_z$ accelerations respectively. We note again that the $m=0$ "phase" is simply an indication of its sign.}
    \label{fig:phase}
\end{figure*}
\begin{figure*}
    \centering
    \includegraphics[width=17cm]{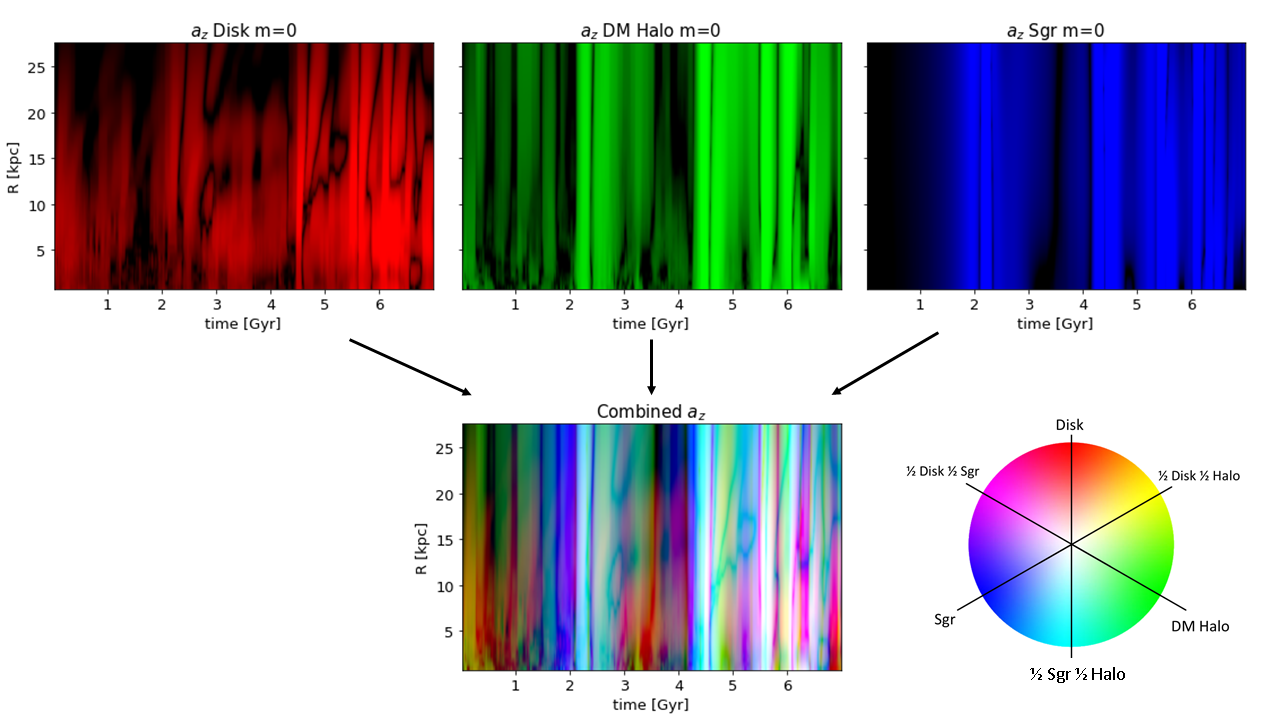}
    \caption{Illustration of how the bottom "RGB" panels in Figures \ref{fig:2d ar} - \ref{fig:2d az} are constructed using the $a_z$ m=0 as an example. Disk self gravity is colored as red, while the tidal accelerations from the DM Halo and Sgr are colored green and blue respectively. In the bottom panel is the resulting combined panel by simply adding these together. Areas with high saturation from multiple components - such as the combined influence from Sgr and the DM Halo at time = 2.5 Gyr - result in a whote hue, while low saturation are black. Red/blue/green colors indicated times and regions where the disk/Sgr/DM halo dominate the acceleration respectively.}
    \label{fig:decomposed rgb}
\end{figure*}

\section{Results II: Analysis of Influences}
\label{sec:results}

We now turn to looking at the net influences on the disk in the context of our understanding of the evolution of Sgr's orbit and mass (Section \ref{sec:sgr}), the (low-level) of global redistribution of matter in all components (Figure \ref{fig:mass distribution}) and the characteristics of the disk response (Section \ref{sec:response}). We first examine the disks own self-gravity (Section \ref{sec:disk}), before comparing the accelerations due to all the Galactic components (Section \ref{sec:components}). We summarize the results in the different temporal Epochs and spatial Regimes in Section \ref{sec:summary}.

\subsection{Disk self-gravity}
\label{sec:disk}

Figure \ref{fig:phase} shows nine pairs of panels to summarize the amplitude (upper panel of each pair) and phase (or sign for the $m=0$, lower panel of each pair) of the Fourier decomposition of the acceleration in the galactic plane due to disk particles. The columns show the different dimensions, $\Delta a_R, a_\phi, a_z $, and the rows separate out the terms, $m = 0, 1, 2$.

\subsubsection{Accelerations within the disk plane}

The in-plane accelerations, $\Delta a_R$ and $a_\phi$, can be directly related to the evolution of surface density apparent in Figure \ref{fig:Fourier density} and discussed in Section \ref{sec:response}. All panels show clear association of sudden changes to the acceleration fields at disk passages.
\\ The \underline{$m=0$ terms} amplitude and sign reflects the outward propagating rings in density during the early part of the simulation. Towards the end, it is dominated by the net change in mass distribution which results in a decrease/increase (red/blue) in the attractive force  in the inner/outer disk. 
\\ The \underline{$m=1$ term} follows the slow, retrograde pattern seen in density in the inner disk for much of the simulation. Since much of the mass is in the inner disk, the accelerations are also dominated by this sense of rotation in the outer disk, in contrast to the prograde pattern seen in density.There is an enhancement in the $m=1$ amplitude in the inner disk (< 5kpc) between the 5th and 6th passages, right before the bar is formed.
\\ The \underline{$m=2$ term} has a prograde pattern at all radii, with new generations of spiral arms excited at each passage and subsequently winding up. The accelerations in this term are most strongly excited by the passages in regions that are localized in radii and that drift inward with each passage. The amplitude following each passage damps more quickly (in less than a Gyr) than the lower order terms. The inner few kpc are more clearly dominated by aligned accelerations throughout the simulation, with power increasing, and pattern speed falling in the last 1.5 Gyrs as the bar forms. 

\begin{figure*}
    \centering
    \includegraphics[width=17cm]{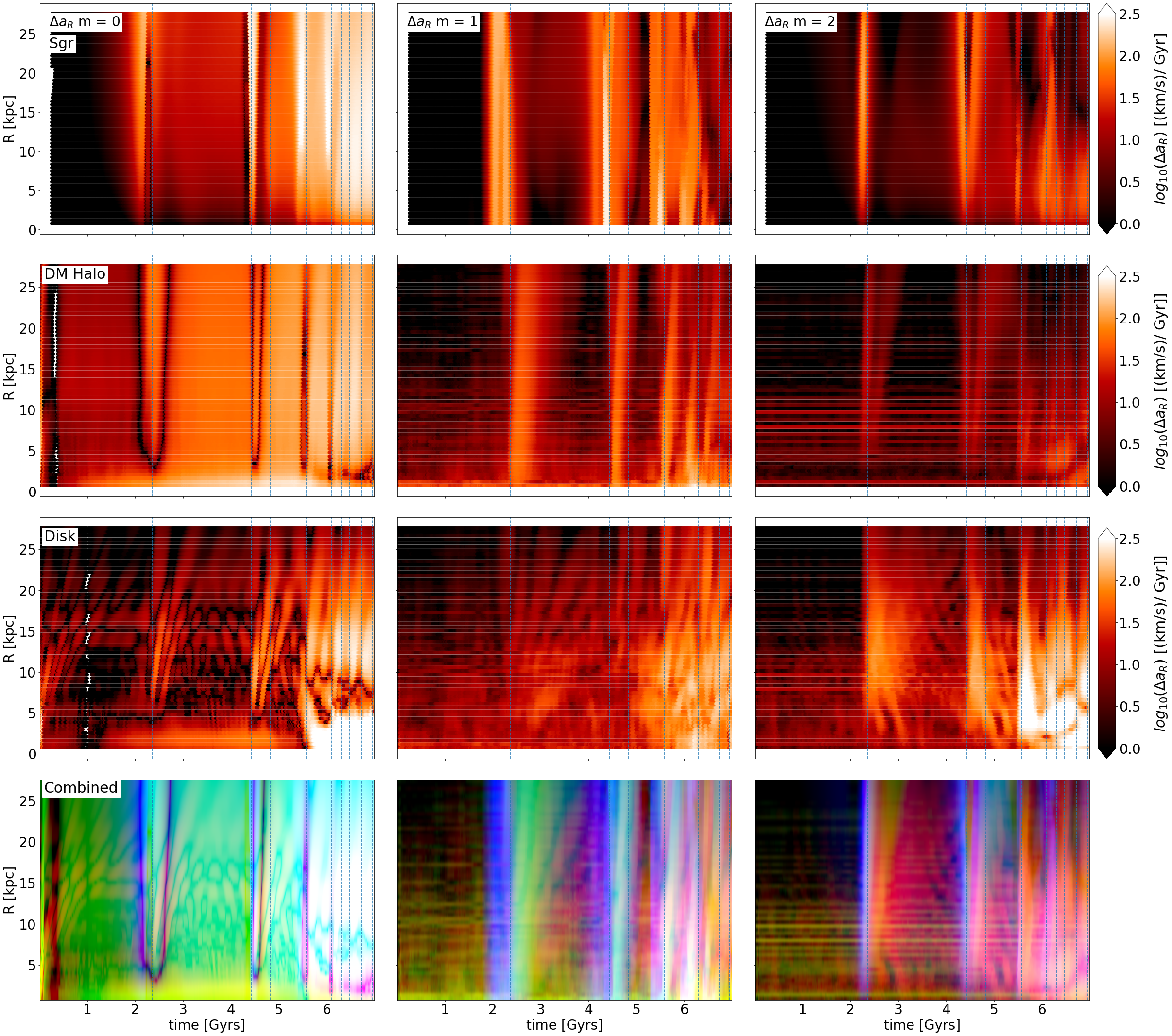}
    \caption{
    Fourier decomposition for the radial acceleration from each component on the disk plane. The x and y axis for each subplot are time (Gyrs) and cylindrical radius away from galactic center (kpc) respectively. The columns shows the amplitude of the m=0/1/2 terms with the vertical dashed lines indicating Sgr disk crossings. Rows in the figure show accelerations in the disk plane due to Sgr (top), the DM Halo (second) and the disk (third). The bottom row shows the combined influences of all three (as in Figure \ref{fig:decomposed rgb}. 
    }
    \label{fig:2d ar}
\end{figure*}
\begin{figure*}
    \centering
    \includegraphics[width=17cm]{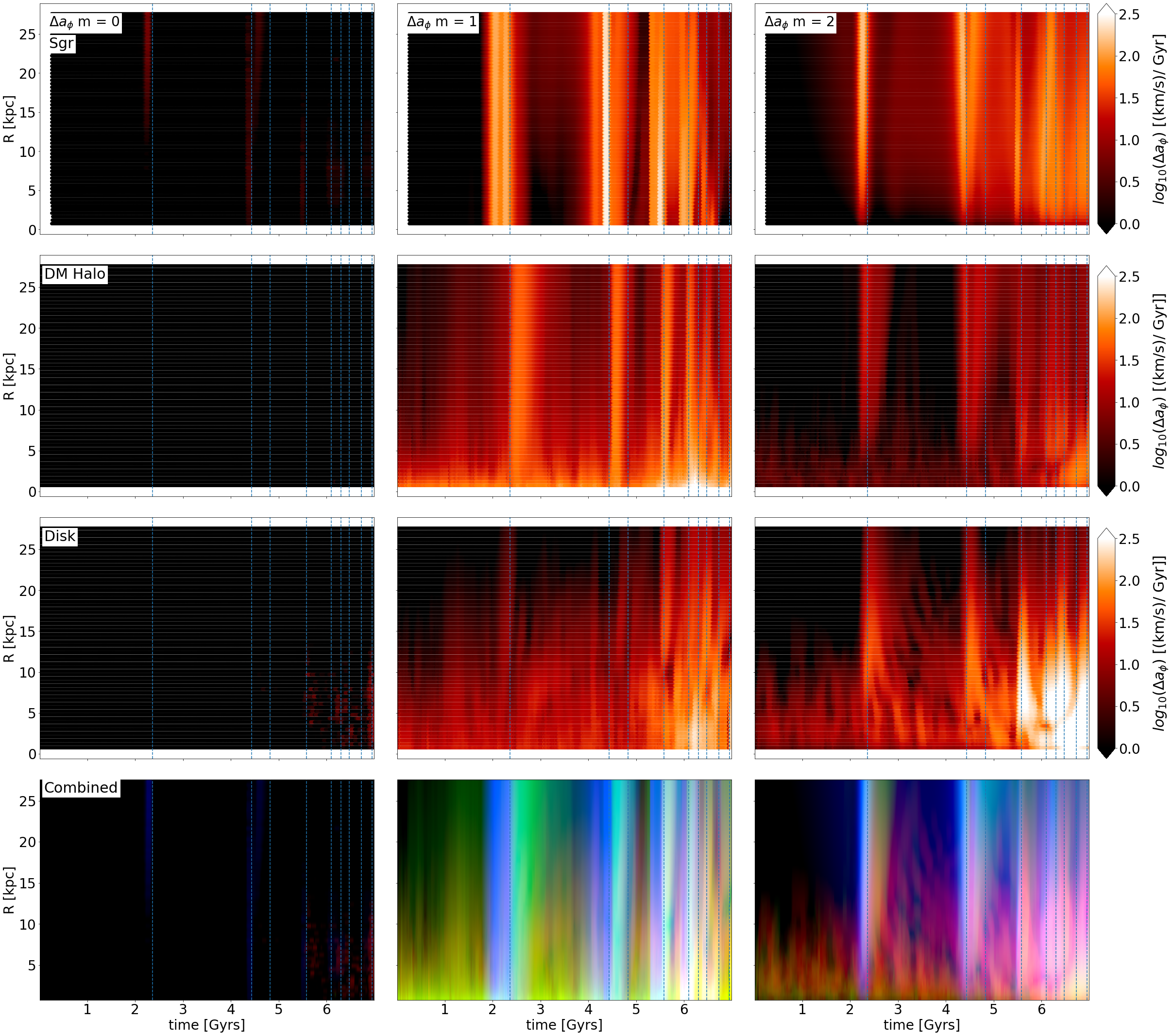}
    \caption{Fourier decomposition of the azimuthal acceleration from each simulation component on the disk plane. Setup the same as Figure \ref{fig:2d ar}. The m=0 column has an amplitude approximately equal to zero because the m=0 mode is, by definition, axisymmetric and so will have no signal in the azimuthal acceleration.}
    \label{fig:2d aphi}
\end{figure*}
\begin{figure*}
    \centering
    \includegraphics[width=17cm]{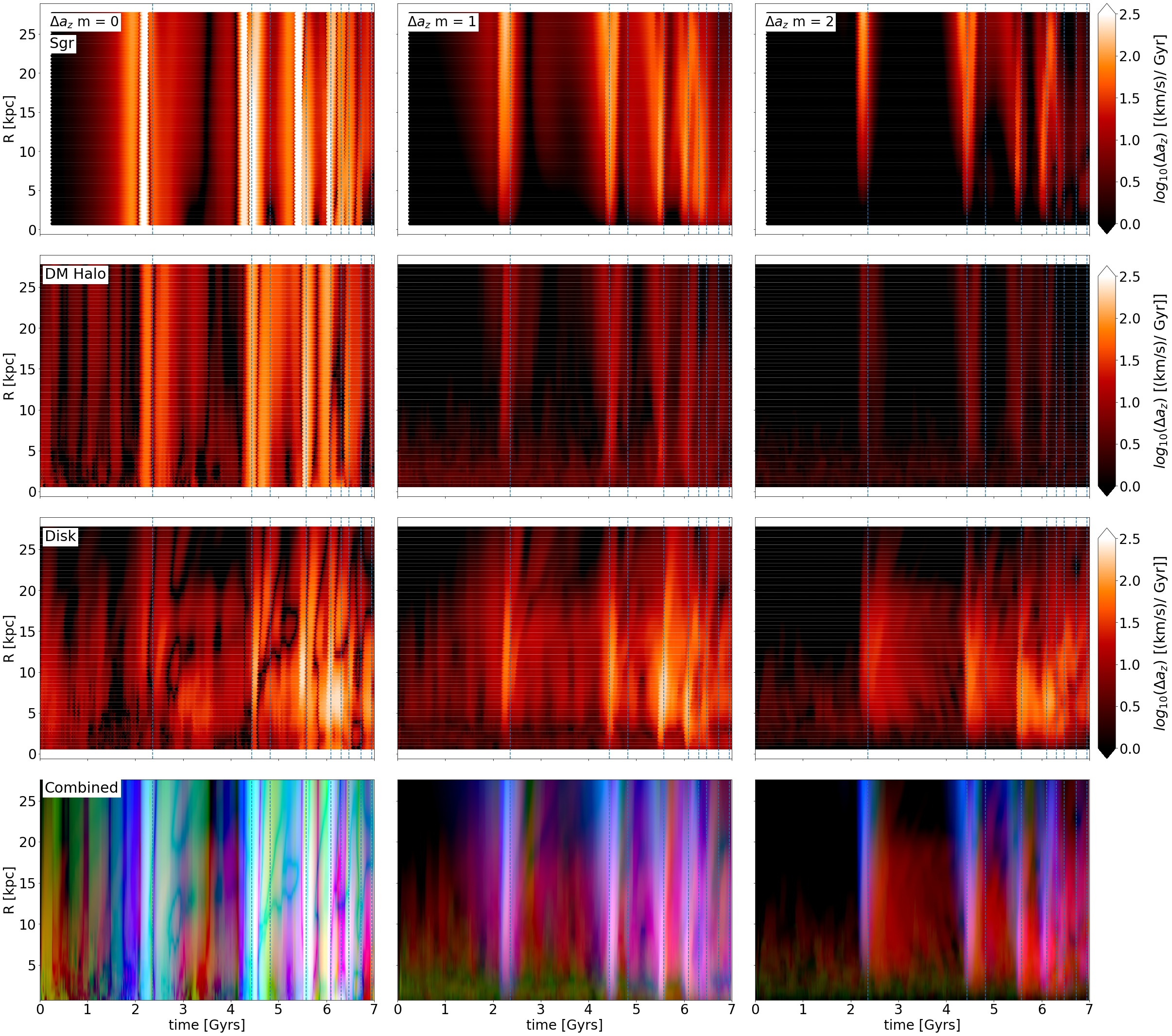}
    \caption{Fourier decomposition of the vertical acceleration from each simulation component on the disk plane. Setup the same as Figure \ref{fig:2d ar}.}
    \label{fig:2d az}
\end{figure*}

\subsubsection{Accelerations perpendicular to the disk plane}

We can directly relate the right hand panels for the acceleration perpendicular to the plane $a_z$, to the decomposition of the vertical structure of the disk of the same simulation by \cite{poggio20} and \cite{laporte18a}. They discussed the structures they found in terms of bending waves, a warp and the formation of a bar. The amplitudes in all three terms are strongly excited for a short period following each disk passage in a localized region that drifts inward in radius over time --- reminiscent of the $m=2$, in-plane accelerations.
\\ The \underline{ $m=0$ term} shows the bending modes in the outer disk that have been studied in great depth elsewhere \citep{widrow14}, clearly excited by each Sgr disk crossing. 
\\ In the \underline{$m=1$ term}, there are signatures of the evolving and outwardly moving warp between 2-4 Gyrs, which is slowly precessing in a retrograde direction \citep[see][for a detailed discussion]{poggio20}.
Note that, unlike the surface density, the vertical oscillations further in remain aligned in phase, indicating the same frequency for a broad range of radii.
Between the 4th and 5th disk crossings there is also power localized in the inner parts ($\sim 5$kpc) of the disk, suggestive of an instability which we will return to in the following sections.
\\ In the \underline{$m=2$ term}, we see the repeated pattern of the passage of Sgr inducing vertical waves, initially aligned in phase,  that subsequently decay, and wind up over time. The $m=2$ motion is distinct  from the $m=1$  warp as the phase-plots indicate that the pattern is prograde rather than retrograde  and rotates at a much higher speed.
A comparison of the phase-plot (bottom right) with the equivalent panel for $a_R$  (top left) suggests that the  timescale for winding the pattern in $Z$ is slightly longer (white lines spaced further apart). This is likely because the in-plane natural phase-mixing frequency ($= \Omega - \kappa/2$) is higher than that for vertical oscillations \citep[$= \Omega - \nu/2$, see][for an in-depth discussion]{poggio20}.




\subsection{Comparison of disk, satellite and halo accelerations}
\label{sec:components}

Plots equivalent to Figure \ref{fig:phase} which include the phases of the Fourier decompositions of accelerations due to Sgr and the DM halo are presented in Appendix A.
In this section we focus on comparing the amplitudes alone, within (Section \ref{sec:rphiacc}) and perpendicular to the Galactic plane  (Section \ref{sec:zacc}).

Figure \ref{fig:decomposed rgb} illustrates our visualization technique for the example of one dimension of acceleration ($a_z$) and one Fourier term ($m=0$).
Each of the upper panels shows the amplitude of this Fourier term at each radius ($y$-axis) and over time ($x$-axis), covering the entire simulation.
Sgr, the DM halo and the disk are shown using blue, green and red color intensity to represent that amplitude of their influence respectively.
In the lower panel, the colors are combined in proportion to the amplitudes from the separate components, with the saturation indicating the combined amplitude (see color wheel on the bottom right of the figure). We will refer to this as the RGB-panel for all subsequent plots. In the RGB panel, red regions indicate where the disk dominates much more than the other components, while yellow regions indicate where  the disk and DM halo influences are comparable to each other. Areas in which all three components have a high amplitude -- i.e. high saturation in their respective colors --  are close to white. On the other hand, areas where the combined influence is low will lead to darker, less saturated colors that are closer to black( e.g.the outer disk during the first Gyr). While there is an immense amount of information in these plots, our use of them will be confined to broadly assessing the importance of the different components in different Epochs (i.e. time) and Regimes (i.e. space). A summary of the RGB panels of the following figures, which allows the reader to compare them all directly, can be found in the Appendix (\ref{fig:2d dominant comp}).

\subsubsection{Accelerations within the disk plane}
\label{sec:rphiacc}

Figures \ref{fig:2d ar} and \ref{fig:2d aphi} show the Fourier decomposition of the radial and azimuthal influences within the disk plane, with each column corresponding to the $m=0/1/2$ terms. Rows 1/2/3 represent the tidal influences of Sgr, the DM halo, and the disk's self-gravity (repeated from Figure \ref{fig:phase}) respectively, and the bottom row shows the RGB panels.

The broad characteristics of the top two rows in both figures clearly reflect our physical intuition for the intertwined evolution of Sgr and the dark matter halo. Sgr's influence has sharp peaks around each of the plane crossings. The small size of Sgr relative to the disk means that its influence is limited in radius and azimuth, and all three Fourier terms are needed to capture this localization. The speed of its passage limits its peak influence to $\sim \pm 100$ Myrs around each plane crossing. The third plane crossing just before $t=5$ Gyrs is unimportant as it occurs near Sgr's apocenter far beyond the edge of the disk.

The dark matter halo responds more ponderously, with the overall contraction and expansion as Sgr dives in and out of its central regions captured in the $m=0$ term of the in-plane radial acceleration. (The sign changes accompanying this can be seen in the $m=0$ panels in Figure \ref{fig:2d ar}). The distortion of the dark matter halo due to Sgr is predominantly captured in the $m=1$ term, with little contribution from $m=2$. This distortion lags the disk crossing, and effectively extends the non-axisymmetric external perturbations to the disk from 100 Myr to $\sim$Gyr timescales. Both these external perturbations move deeper into the disk over time as Sgr's orbit decays due to the energy and angular momentum exchange between the components.

In contrast to Sgr and the DM halo, the disk's distribution of near-circular orbits supports coherent, rapidly rotating and mixing responses seen in density, and their corresponding accelerations. These are clearly instigated by each passage and decay over tens of orbital periods, as noted in the prior section.

The bottom RGB panel summarizes the relative importance of each of these contributors. The blue hue from Sgr is clearly visible at every disk passage in every panel of both figures. The importance of the distortion of DM halo in enhancing the torque on the disk is apparent as extended green colours in the $m=1$ panels following each Sgr disk crossing \citep[as recognized in many prior works][]{weinberg98, weinberg06,gomez13,laporte18a}. The increasing relevance of $m=2$ patterns in the disk --- presumably the spiral arms and the bar seen in Figure \ref{fig:regimes} --- can be seen in red, in particular in the bottom right panel.

\subsubsection{Accelerations perpendicular to the disk plane}
\label{sec:zacc}

Figure \ref{fig:2d az} repeats Figures \ref{fig:2d ar} and \ref{fig:2d aphi} for the $a_z$ component of accelerations due to each galactic component.

As with the earlier figures, Sgr's influence (top row) is confined to a narrow window around each of the disk crossings and is apparent in each Fourier term.
The characteristics of the left column can be simply related to Sgr's motion, as it orbits perpendicular to the disk in a plane almost aligned with the $x$-axis. The abrupt zeros in amplitude (black) in every panel reflect this geometry as Sgr plunges in sequence down and up through the $z=0$ plane. (This can also be seen explicitly in the associated phase-plots in Figure \ref{fig:sag phase}.)  Over time, the influence becomes ever more localized in radius as Sgr's orbit decays and the crossings occur within the disk itself.

In contrast to Sgr, the DM halo's influence perpendicular to the plane (second row of paired panels)  is largely captured by the $m=0$ term alone as it responds  to the much smaller intruder, and pulls vertically on the galactic disk. This response noticeable lags the crossing time. Nor is it strictly confined to the disk passages, but rather takes $\sim 1$ Gyr to settle following each encounter. The much lower power in the higher order terms suggests that the DM halo does not tilt significantly relative to the disk despite Sgr's influence.

\begin{figure*}
    \centering
    \includegraphics[width=17cm]{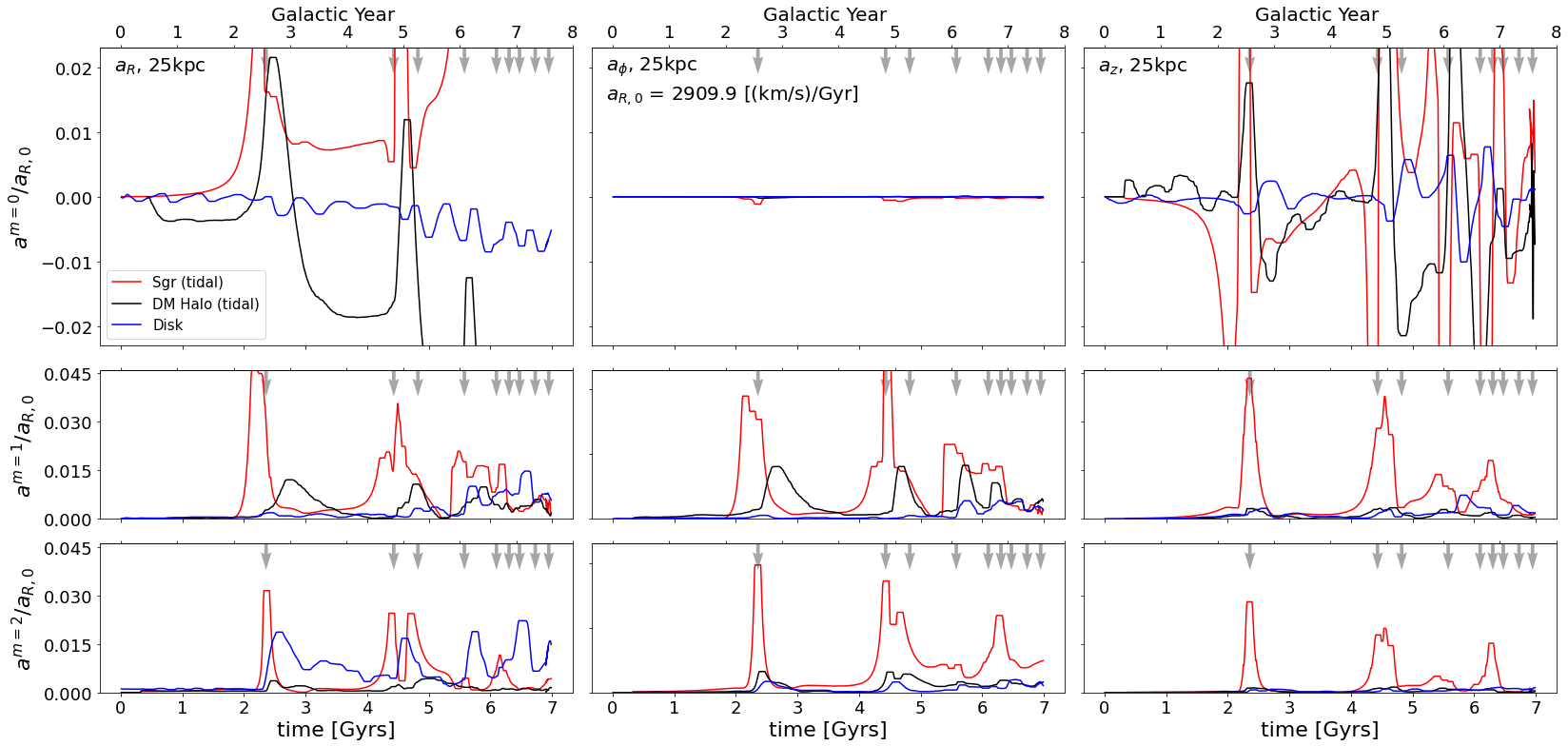}
    \caption{Fourier amplitudes for each cylindrical acceleration as a function of time. Amplitudes shown here are only for particles at a galactocentric radius of 25 kpc away from the center. The figure shows the first three Fourier modes for each acceleration term [$a_R$ in the first column, $a_{\phi}$ in the second, and $a_z$ in the last]. The modes are each normalized by the radial m=0 value (summed from the DM Halo and Disk) of the corresponding ring at time = 1 Gyr. The normalization factor, $a_{R,0}$, is printed in the top middle panel. Each simulation component's Fourier amplitude is represented by a different color: Sgr is colored as red, the DM halo as black, and the Disk as blue. Time is shown on the x axis both in Gyrs (lower x axis) and in galactic years (upper x axis) for that particular radius - in this case 25 kpc. Arrows on the top of each panel show the time which a passage with Sgr occurs.}
    \label{fig:line amp 25}
\end{figure*}

\begin{figure*}
    \centering
    \includegraphics[width=17cm]{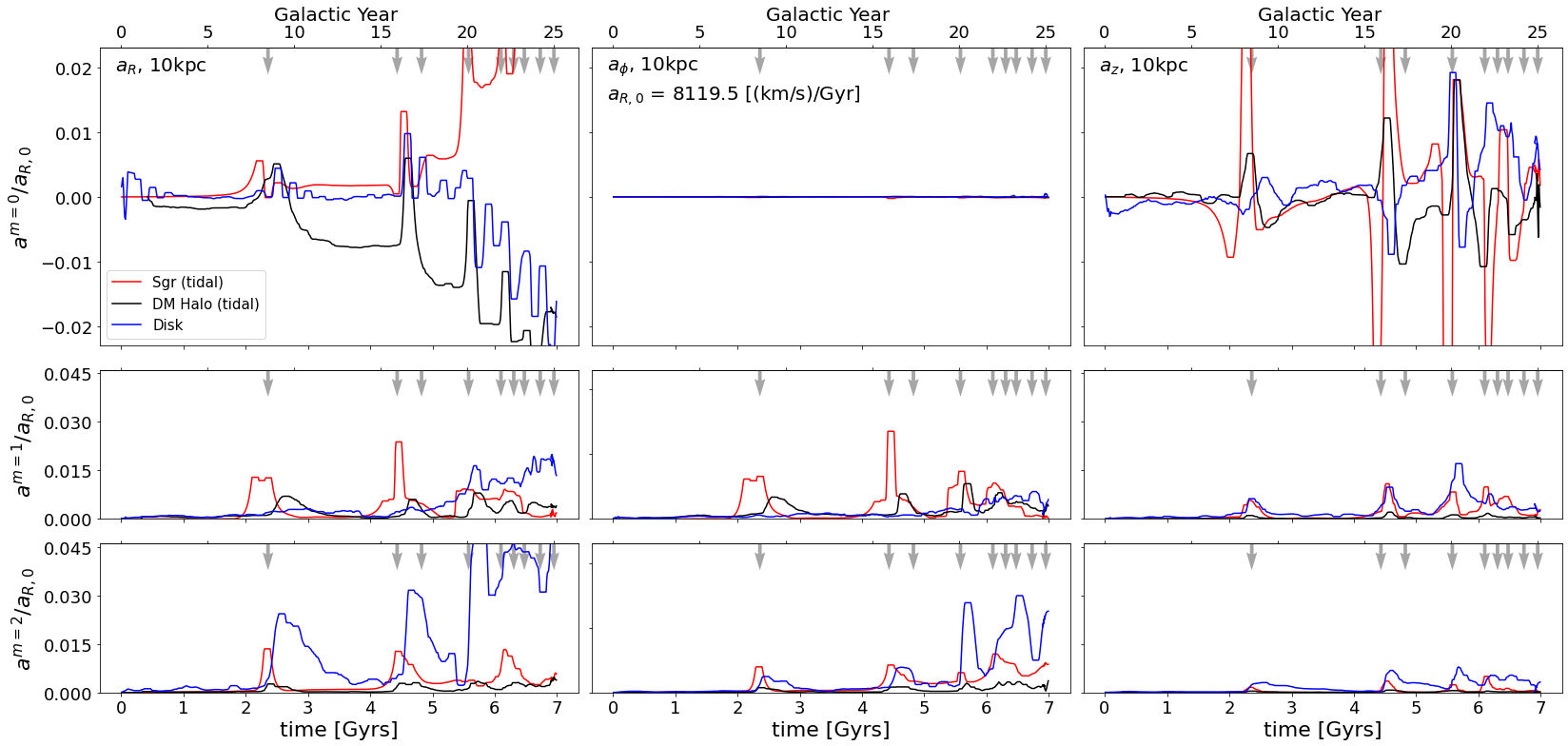}
    \caption{Fourier amplitudes at a constant radius as a function of time for particles 10 kpc away from the galactic center. Layout exactly the same as Figure \ref{fig:line amp 25}}
    \label{fig:line amp 10}
\end{figure*}

\begin{figure*}
    \centering
    \includegraphics[width=17cm]{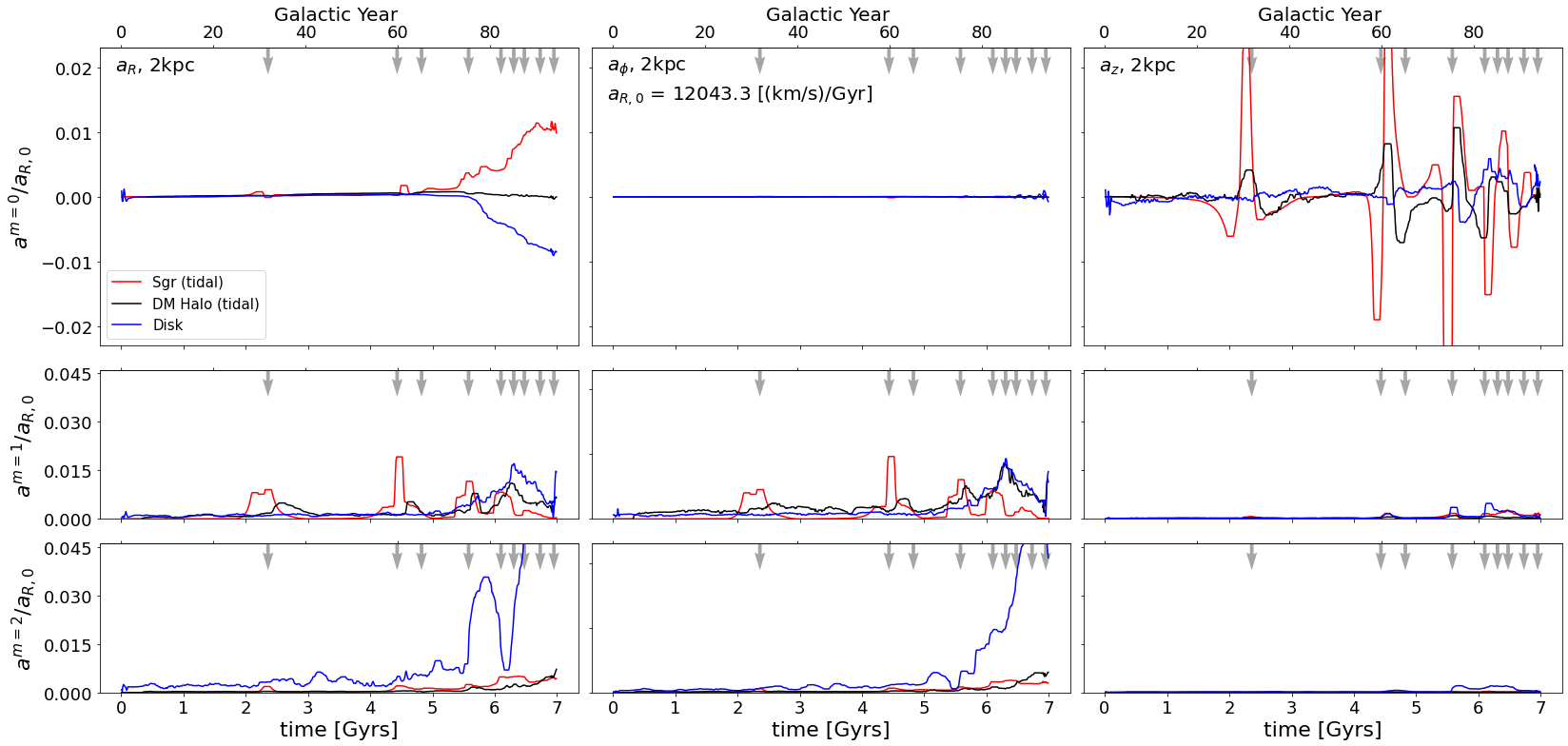}
    \caption{Fourier amplitudes at a constant radius as a function of time for particles 2 kpc away from the galactic center. Layout exactly the same as Figures \ref{fig:line amp 25} and \ref{fig:line amp 10}}
    \label{fig:line amp 2}
\end{figure*}

Nevertheless, the combination of Sgr and the DM Halo does instigate bending in the disk. This is captured by localized power in both the $m=1$ and $m=2$ terms (third row of panels), coincident with Sgr's passages. The self-gravity of this bending is far more important than the DM halo's external influence in these terms. It extends the influence of vertical forces beyond Sgr's disk passages, and exceeds the amplitude of those impulsive perturbations by the end of the simulation.

The RGB-panel demonstrates that in the $z$ direction, the DM halo's response to Sgr (green) is most apparent in the $m=0$ term, effectively extending the duration of net vertical accelerations beyond each crossing. The self-gravitating influence of bending of the disk (red) is important in the $m=1$ and $m=2$  terms only for a short time following earlier passages, but becomes increasingly relevant as Sgr's orbit decays. In particular, note
the red hues around $R\sim$5 kpc in $m=1$ which suggest significant instability in the inner disk.
Sgr's influence around the disk crossings for these higher order terms is indicated by the violet hues at those times.

\subsection{Comparison of influences in temporal epochs and spatial regimes.}
\label{sec:summary}

Figures \ref{fig:2d ar} - \ref{fig:2d az} provide a visual overview of how each component influences the disk, in what region, over what timescale and in which Fourier term.  
As a quantitative comparison of the significance of these different influences on disk, Figures \ref{fig:line amp 25} - \ref{fig:line amp 2} show the amplitude of Fourier terms in the outer (25 kpc), middle (10 kpc) and inner (2 kpc) disk respectively. 
The columns show the three cylindrical acceleration components ($a_z$, $a_{\phi}$, $a_R$) and the rows are the first three Fourier terms ($m=0$, $m=1$, $m=2$). The values are presented as fractions of the radial $m=0$ acceleration of the corresponding ring at time = 1 Gyr. Each component's amplitude is plotted together in order to compare their relative strength as a function of time and radius (Sgr in red, the DM halo in black and the disk in blue). As before, for the $m=0$ terms in these figures we plot the change in amplitude with respect to the conditions at time = 1 Gyr. 

Figure \ref{fig:line amp 25}, showing accelerations in the outer Galaxy, emphasizes how this region of the disk is sculpted by the twin influences of Sgr and the DM halo. The mass in disk stars is sufficiently low that its self-gravity is largely negligible in this region until its center becomes non-axisymmetric towards the end of the simulation. In the $z$-direction, Sgr is dominant in all Fourier terms. As our expansion is centered on the plane defined by the disk stars, we can see how the offset of the disk from the DM halo's symmetry plane is apparent as an acceleration in the $z$ direction in the apposite direction to Sgr, lagging behind the passages. The non-axisymmetric distortion of the DM halo (the ``wake'') induced by Sgr contributes significant amplitude in the $m>0$ terms for both $a_\phi$ and $a_R$ that extends beyond the influence of Sgr alone at the disk crossings, as anticipated and demonstrated from prior work \citep{weinberg98,weinberg06,gomez16,laporte18b}. The steady decrease in the $m=0$ DM halo contribution  stems from the center of the halo being heated by the combined interactions and becoming more diffuse.

Figure \ref{fig:line amp 10} repeats Figure \ref{fig:line amp 25} but for a ring 10 kpc away from the galactic center. In this region, Sgr, the disk and the DM halo all exert significant influences. In the $z$-direction, the amplitude of the disk oscillations reveal the presence of bending modes (in $m=0$), a warp (in $m=1$, coincident with the DM halo response) and a phase-mixing $m=2$ vertical oscillation. In the azimuthal direction the extended DM halo torque is again apparent in $m=1$, while the formation of spiral arms can be seen in $m=2$. In the radial direction, the (mild) expansion of the middle parts of both the DM halo and disk are apparent in $m=0$ (see Figure \ref{fig:mass distribution}), as well as the decrease in the Galactocentric radius of Sgr. The $m=1$ and $m=2$ terms for the disk suggest the increasing importance of a bar as it is increasingly destabilized by repeated Sgr passages.

Figure \ref{fig:line amp 2} repeats Figures \ref{fig:line amp 25} and \ref{fig:line amp 10}
for a ring in the inner disk, at 2 kpc from the galactic center. In this region, while the evolution is driven by the Sgr impacts, disk self-gravity dominates the acceleration fields.
In particular, the spike in the amplitude of the $a_z$, $m=1$ term in the disk following Sgr's disk passage at $t\sim 5.6$Gyr suggests that this is the disk instability event that led to the formation of the Galactic bar whose presence is clear following this passage in the $m=1$ and $m=2$ terms in $a_\phi$ and $a_R$
\citep[see][for an in depth description of this process]{petersen19,weinberg20}.

\begin{figure*}
    \centering
    \includegraphics[width=17cm]{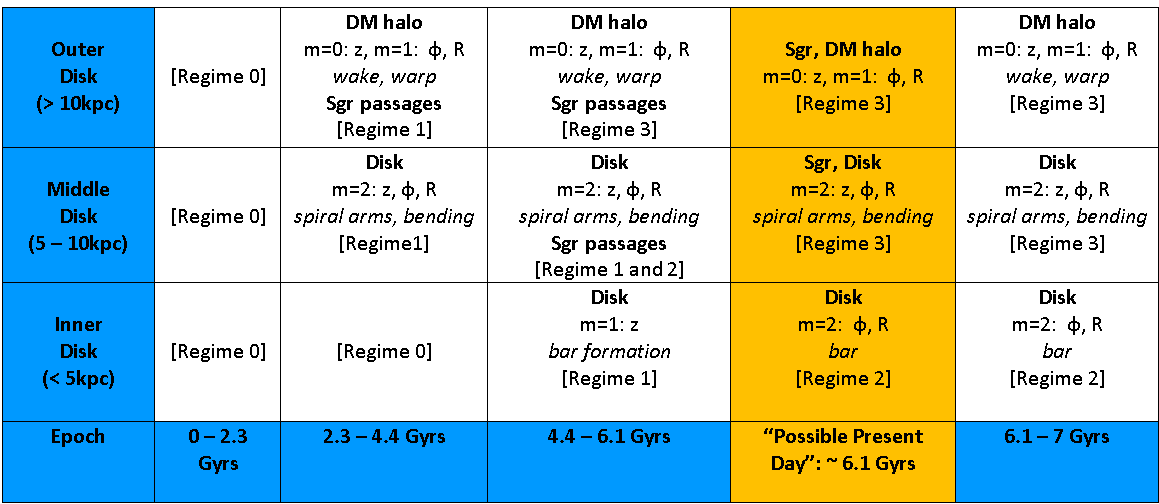}
    \caption{Summary of dominant perturbative influences. Bold font indicates source, regular font indicates Fourier term and dimension in each temporal epoch.  Italics indicate physical consequences of the disturbances. Interaction Regimes defined in Section 3 are in square brackets: the disk evolving in isolation [0]; after a single impact [1]; during impacts repeated roughly every orbit [2]; during multiple impacts per orbit [3]. }
    \label{fig:end}
\end{figure*}

\section{Summary and Conclusion}
Many prior studies have shown that it is impossible to understand the response of a galactic disk to the gravitational influence of a satellite galaxy without taking a holistic view — one that accounts for the mutual interactions of all galactic components \citep{weinberg98,weinberg99,vesperini01, weinberg06,gomez16, laporte18a, laporte18b, garavito20a}.
The passage of the satellite distorts the DM Halo surrounding the disk ; this distortion enhances the influence on the disk both directly and by causing the orbit of the satellite to decay; and the disk’s own self-gravity can drive its further evolution.
Even the Milky Way - a relative quiet backwater in the Universe - shows abundant signatures of these ongoing actions and reactions \citep{widrow12,williams13,amores2017,antoja18,binney18,laporte19, bland-hawthorn19,chen2019,skowron2019, laporte2021}, which could potentially tell us much about how the components are interacting. However, disentangling the interlocked causes of these signatures remains a challenge to the field. 

In this work, we have presented a new approach to studying this complexity. We mapped accelerations caused by the three main protagonists (Sgr, Galactic disk, DM halo) onto the Galactic disk in the radial, azimuthal and vertical directions. The mapping allows us to determine the impact of each player (and therefore which one is dominant) at a given position in the Galactic disk and at a given time throughout the simulation, providing a unique view on the causal connections driving the evolution of the Sgr-MW system. 
We have shown how each different component's influence on the disk can lead to the asymmetries present in our galaxy and how some signatures are more heavily dominated by one component over another. 

Figure \ref{fig:end} presents a schematic to summarize our results. It tabulates the nature of the dominant influences (source, Fourier term and dimension), response (in italics) and interaction regime (square brackets) for different disk regions (rows) and within each  temporal epoch identified in Section \ref{sec:results}. For instance, during 3rd epoch (4.4 - 6.1 Gyrs), we have three different interaction regimes depending on which spatial region of the disk we analyse. 

The figure emphasizes that each component - Sgr, the DM halo and the disk -  shapes the disk in distinct and characteristic ways in space and time. 
\begin{itemize}
    \item {\bf Sagittarius} directly dominates accelerations only for $\pm 100 - 300$ Myears (depending on radius) around plane crossings at small enough radii to impact the disk, exciting longer-lived, slowly decaying modes in the disk. Its direct influence is particularly important during the first few encounters while it still has a mass $\sim 2-6 \times 10^{10} M_\odot$, and in the intermediate-outer disk where the encounter timescale is longer.
    \item{\bf The Dark Matter Halo distortion} (``wake'') enhances and extends the direct influence of the much smaller dwarf, both in $m=0$ vertical acceleration and as $m=1$ in plane accelerations. These distortions are the dominant influence on the outer disk, driving the formation and evolution of the warp.
    \item{\bf The self-gravity of the disk response} becomes increasingly important at smaller radii and as the interaction progresses. The scale of accelerations due to bending waves in the middle-disk rivals the importance of the initial disturbance itself. The inner disk clearly becomes unstable in $m=1$ vertical oscillations that lead to the formation of a bar ($m=2$ terms, aligned in phase) by 6Gyrs.
\end{itemize}

Given the similarity of the simulations to the properties of the observed Milky Way and Sgr, it is likely that this description represents the broad characteristics of the recent dynamical history of our own Galaxy. Of course, the details are not expected to exactly match for a multitude of reasons. For example, the extent of the influence of Sgr will depend on the details of its mass loss history and  we have not included the current influence of the Large Magellanic Cloud \citep[expected to contribute equally to the outer disk, see][]{laporte18a,laporte18b}. In addition, our simulation assumed non-evolving disk and DM halo components, initially in equilibrium, while we might expect the Milky Way to grow in scale and mass over several Gyrs and only just be recovering from its merger with Gaia-Sausage-Encedaelus \citep[see][]{helmi18,myeong19,koppelman19,lancaster19,grand20}. This missing history in turn leads to a disparity between the suspected age of the Milky Way's bar which is thought to be many Gyrs old \citep{debattista19,bovy19} and the recent formation of the  bar  in the simulation, within the last Gyr.

Nevertheless, it is interesting to speculate on the implications of our work for our view of the Milky Way today, as summarized by the orange column in Figure \ref{fig:end}, which corresponds to the 5th plane crossing in the simulation after $\sim$6 Gyrs of evolution.  
The inner disk is dominated by self-gravitating structures (e.g. the bar), likely built from its (2nd order) reaction to its own (1st order) response to prior interactions.
Prospects for dissecting  the dynamics of the ongoing interaction are more promising at larger radii.
In the middle region, with Sgr currently plunging through the disk of the Milky Way \citep[with an imminent plane-crossing in $\sim 25$Myrs --- see][for further discussion]{gandhi2021} we are entering a period which coincides with the peak of its influence and we most plausibly might catch the signature of the ongoing interaction directly. One approach could be to look at a global distortion to the disk's velocity field \citep[see][]{laporte19} before it has time to phase-mix away, although any interpretation will necessarily be complicated by the disk's own influence. Of course, the amplitude depends on the mass still associated with the dwarf. In our simulations, the mass of the dwarf at this point is sufficiently low that the disk's own self-gravity is more important, and recent estimates for the mass of the dwarf suggest it could be as low as \citep[e.g. $5\times 10^8 M_\odot$][]{vasiliev20}. 
It is clear that the outer disk \citep[e.g.][]{laporte2021} is the place to search for clues about the DM halo's distortion and previous impacts, though in reality the influence of both Sgr and the LMC will need to be taken into account in any interpretation \citep{laporte18b}.

We conclude that our holistic review of the interaction itself motivates the next step: a holistic review of the signatures of that interaction.

\vskip 0.3in
\section*{Acknowledgements}

We would like to thank the Center for Computational Astrophysics in the Flatiron Institute for allowing us access to their binder server, making the management of the simulations much easier to share among the other members of this collaboration. We would also like to thank the members of both the Milky Way Stars group at Columbia and the Dynamics Group at the CCA, who consistently provided useful comments and suggestions for improvement of this work. In particular, we would like to thank Adrian Price-Whelan for greatly assisting in the creation of figure \ref{fig:2d dominant comp}. We also acknowledge fruitful discussions with Martin Weinberg and  Team Beefy. KVJ and DGF were supported by NSF Grant AST-1715582. CL acknowledges funding from the European Research Council (ERC) under the European Union’s Horizon 2020 research and innovation programme (grant agreement No. 852839). This work was supported in part by World Premier International Research Center Initiative (WPI Initiative), MEXT,Japan. This work used the Extreme Science
and Engineering Discovery Environment (XSEDE) through the grant TG-AST150025, which is supported by National Science Foundation grant number OCI-1053575.

\vskip 0.3in
\section*{Data Availability}
\textit{The data underlying this article were accessed from the binder cluster at the Computational Center for Astrophysics (CCA) at the Flatiron Institute. The simulated data used in this research will be shared on reasonable request to the corresponding authors.}

\bibliographystyle{mnras}
\bibliography{export-bibtex} 


\appendix

\section{Summary Plots for Disk Influences}

\begin{figure*}
    \centering
    \includegraphics[width=17.5cm]{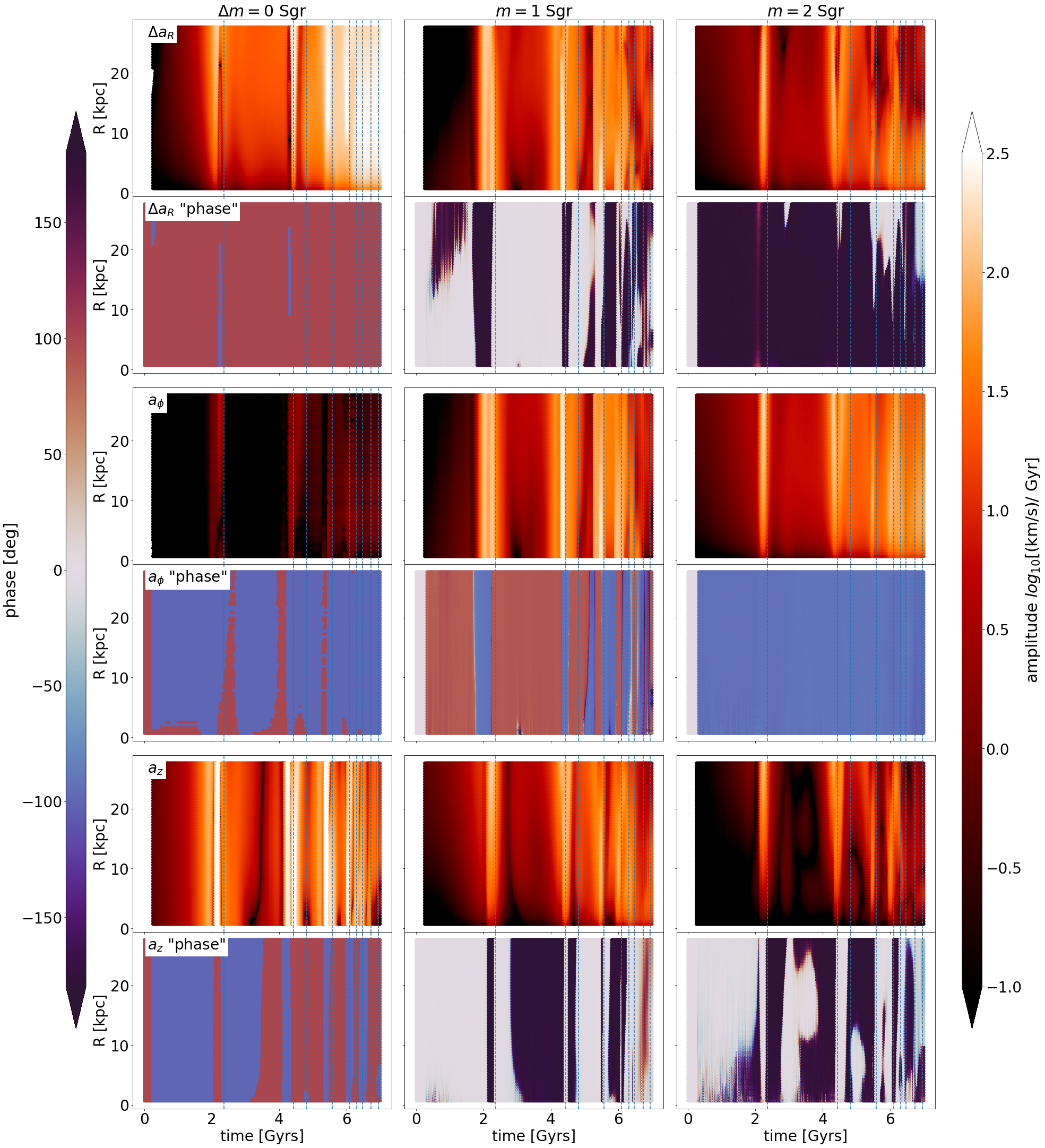}
    \caption{Amplitudes and Phases for Sagittarius. Layout is the same as Figures \ref{fig:phase} and \ref{fig:dm halo phase}}
    \label{fig:sag phase}
\end{figure*}

\begin{figure*}
    \centering
    \includegraphics[width=17.5cm]{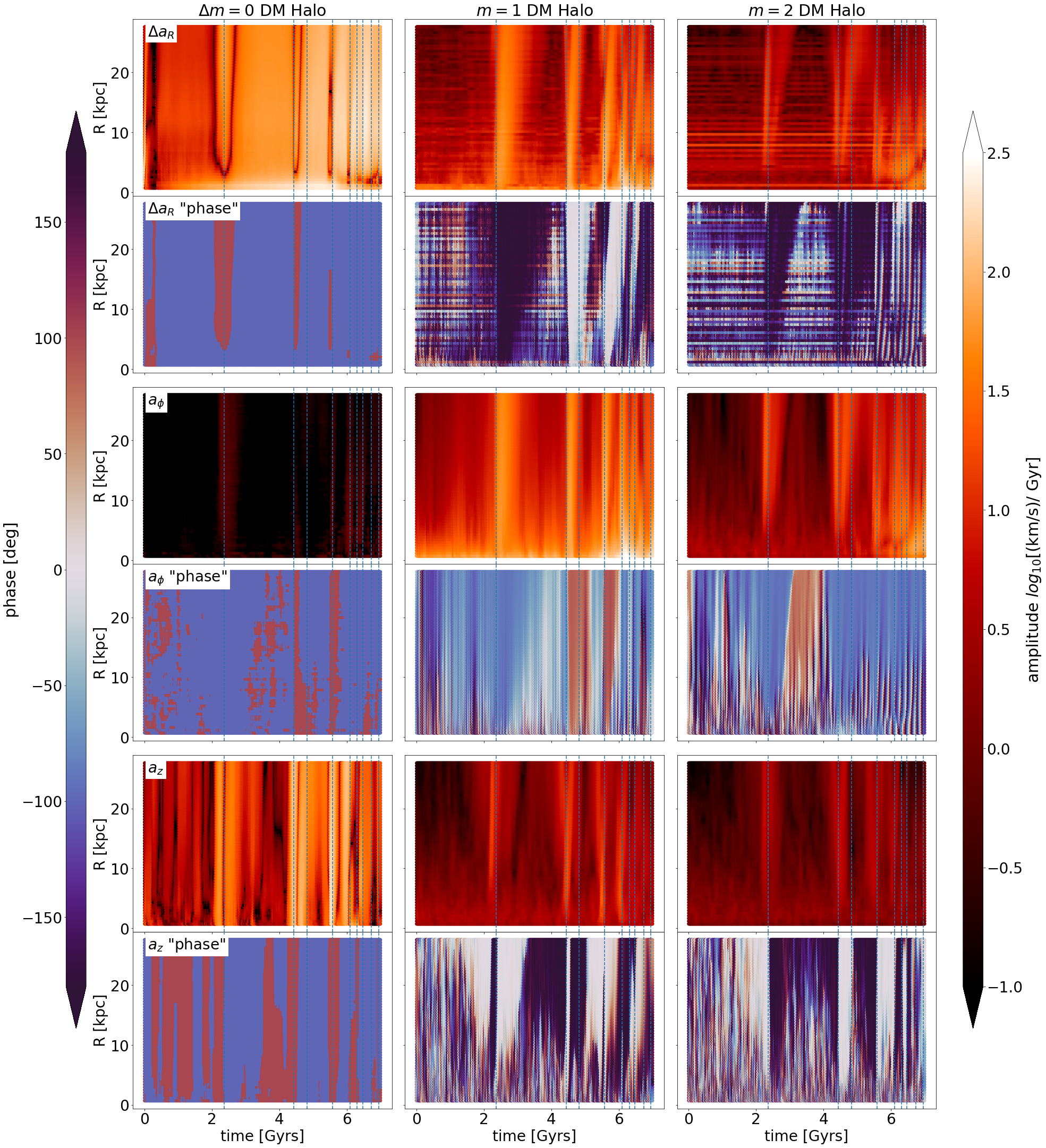}
    \caption{Amplitudes and Phases for the Dark Matter Halo. Layout is the same as Figures \ref{fig:phase} and \ref{fig:sag phase}}
    \label{fig:dm halo phase}
\end{figure*}

\begin{figure*}
    \centering
    \includegraphics[width=17.5cm]{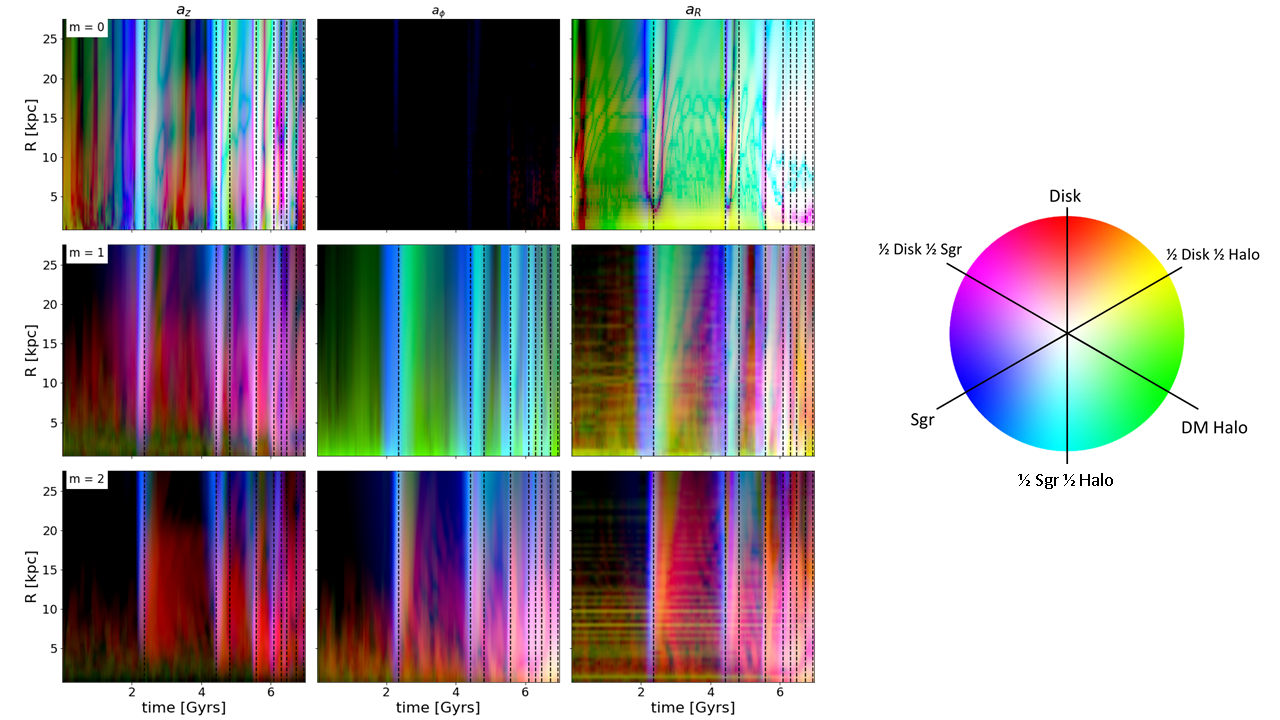}
    \caption{Fourier decomposition and comparison of the accelerations on the disk midplane due to each simulation component. Each panel shows the Fourier amplitude as a function of galactocentric radius $R$ (y-axis) and time (x-axis) from all components. The columns are organized by acceleration, going from left to right with $a_z$, $a_{\phi}$, and $a_R$. Rows are organized by ascending Fourier modes. For each Fourier mode and cylindrical acceleration, the component's respective influence on the disk is colored separately and compared. The tidal influences from the DM Halo and Sgr are colored in green and blue respectively, while the disk self-gravity is colored in red. Areas where more than one component have a large influence get a mix of these three colors. The higher the amplitude of a given component's influence, the more saturated the respective color will be. As such, areas where all three components have high saturation, the combination of all colors will be close to white (an example of this is the right side of the top-right panel). On the other hand, low amplitudes reduces the saturation, meaning that darker regions have low influence (outer disk during the first Gyr in the bottom left panel). 
    To the right of the figure is a color wheel legend for how to interpret this mix of colors. An example of how a panel from this figure is assembled can be found in Figure \ref{fig:decomposed rgb}}
    \label{fig:2d dominant comp}
\end{figure*}

\bsp	
\label{lastpage}
\end{document}